\documentclass[aip,jap,reprint,nofootinbib,longbibliography]{revtex4-2}

\usepackage[T1]{fontenc}
\usepackage[utf8]{inputenc}
\usepackage{booktabs}
\usepackage{bm}
\usepackage{amsmath}
\usepackage{amssymb}
\usepackage{graphicx}
\usepackage{siunitx}
\usepackage[dvipsnames]{xcolor}
\usepackage{soul}   

\usepackage{hyperref}

\newcommand{\SAR}{\mathrm{SAR}}
\newcommand{\ta}{\tau_{\mathrm{N}}}
\newcommand{\tsc}{t_{\mathrm{s}}}
\newcommand{\sig}{\sigma}
\newcommand{\om}{\omega}
\newcommand{\chipp}{\chi''}
\newcommand{\chiz}{\chi_0}
\newcommand{\Ha}{H_{\!a}}
\newcommand{\LLL}{\mathrm{LLL}}
\newcommand{\LRT}{\mathrm{LRT}}

\begin{document}

\title{Specific absorption rate of uniaxial single-domain nanomagnets:
stochastic spin dynamics versus linear response theory}

\author{H.~Kachkachi}
\email{hamid.kachkachi@univ-perp.fr}
\affiliation{Laboratoire PROMES, CNRS (UPR~8521) \&
Universit\'e de Perpignan Via Domitia,
Rambla de la Thermodynamique, Tecnosud, 66100 Perpignan, France}

\begin{abstract}
We compute the specific absorption rate of a uniaxial
single-domain nanomagnet driven by an alternating magnetic field by two methods: i) direct numerical integration of the stochastic
(Langevin) Landau--Lifshitz--Gilbert equation (the LLL approach),
and ii) linear response theory (LRT) based on the Debye susceptibility
with the Néel relaxation time $\ta$. We first analytically show
that both methods are equivalent for small magnetic field amplitude, and then
compute their deviation $\Lambda\equiv\SAR_{\LLL}/\SAR_{\LRT}-1$
as a function of the magnetic field amplitude for two temperatures chosen on
opposite sides of the Debye resonance. One of the main results is
that the sign and magnitude of $\Lambda$ are governed by the dimensionless
product $\om\ta$, in addition to the linearity parameter $\xi=\mu_{s}B_{0}/k_{B}T$
for the easy-axis geometry considered here. Indeed, below resonance ($\om\ta<1$),
linear response theory overestimates the specific absorption rate, whereas above resonance ($\om\ta>1$, the
regime typical of blocked nanoparticles) linear response theory can underestimate the
specific absorption rate by up to $\sim70\%$ at $\xi\sim2$. We expect this work to
provide a quantitative guidance for the use of linear response theory in magnetic hyperthermia
and related nanoscale heat-transport problems, and to serve as a single-particle
benchmark for extensions to many-spin and interacting systems.
\end{abstract}


\maketitle

\section{Introduction}

\label{sec:intro} 

Nowadays, heating of magnetic nanoparticles by an alternating magnetic
field (AMF)---magnetic hyperthermia (MH)---has become a serious
cancer-therapy modality sustained by an increasing attention~\cite{Pankhurst2003,Dutz2014,Hergt2006, Zhang2024MIHReview, Baldea2025NanoMH, Bai2024MHTTheranostics}.
More broadly, it is a rich playground for studying heat generation and
transport at the nanoscale~\cite{DejKac2022,Dejardin2024,Iglesias2021,Ruta2024,Kachkachi2026LocalGlobalHeating}.
In this context, the figure of merit is provided by the specific loss
power (SLP), or more often by the specific absorption rate (SAR),
\emph{i.e.}, the power dissipated per unit mass of magnetic material.

In practice, the SAR is very often estimated within the \emph{linear
response theory} (LRT), through the compact expression $\SAR_{\LRT}=\mu_{0}\om\chipp(\om)H_{0}^{2}/(2\rho)$,
where $H_{0}$ is the field amplitude, $\om=2\pi f$ the angular frequency,
and $\chipp(\om)$ the imaginary part of the complex susceptibility~\cite{Rosensweig2002,Hergt2006,Dejardin2017,DejKac2022}.
The advantage of this approach is that, combining the Debye relaxation
model and the Néel--Brown expression for the relaxation time~\cite{Neel1949,Brown1963},
it gives a closed-form prediction that requires no (heavy) numerical
simulations. The obvious question that arises then is when LRT is
quantitatively valid. In general, in the literature, the answer is when the linearity condition $\xi\equiv\mu_{s}\mu_{0}H_{0}/k_{B}T\ll1$
is satisfied~\cite{Carrey2011,Andreu2013}. However, we know that
this criterion is necessary but not sufficient. By way of example,
direct comparisons between stochastic LLG simulations and LRT for
iron-oxide assemblies~\cite{Usov2021}, and between linear and nonlinear
AC-susceptibility methods~\cite{Yamaminami2021}, demonstrate that
the sign and magnitude of LRT errors depend on other parameters, in
addition to $\xi$.
While these studies establish that LRT can fail significantly beyond
the $\xi\ll1$ regime, none provides a closed-form analytical expression for
the nonlinear correction to the SAR, nor identifies the precise boundary in
the $(\sigma,\xi,\eta)$ parameter space at which the sign of the LRT error
reverses. As we show below, this sign reversal occurs not at the Debye
resonance $\eta=1$ but at a larger value $\eta_{c}(\sigma)$, which exceeds unity for $\sigma>3$ and tends to $\sqrt{3}$ at large $\sigma$, a result with
direct practical implications for both fast-relaxing and blocked
nanoparticles.

An alternative, exact in the macrospin approximation, numerical approach
proceeds by integrating the stochastic Landau--Lifshitz--Gilbert
equation (LLGE) directly. In the sequel, this will be referred to
as the Landau--Lifshitz--Langevin (LLL) method. Then, the SAR may
be estimated as the time-averaged hysteresis-loop area~\cite{Lacroix_etal_JAP2009, Carrey2011, martinez2013learning, Mehdaoui_prb2013}.
LLL makes no linearity assumption and is one of the appropriate benchmarks
for LRT. Time-quantified Monte Carlo (TQMC) simulations, in which
a Monte Carlo step corresponds to a physical time unit derived from
the attempt frequency~\cite{ChubykaloEtAl_prb03, nowaketal00prl, Ledue2026}, can play the same role and
are particularly suited to multi-spin systems.

In the present work, we consider the simplest system of a single uniaxial nanomagnet, which is the building block of a nanoscale heating system. This macrospin study is meant as a preliminary study and a benchmark for our ongoing investigation of assemblies of interacting nanomagnets, such as open chains, already considered in Ref.~\onlinecite{Ledue2026}. Then, we study the validity of LRT with respect not only to $\xi$ alone, but also to the dimensionless product $\om\ta$.
Indeed, the Debye susceptibility $\chipp$ peaks at the \emph{resonance temperature} $T_{\mathrm{peak}}$ where $\om\ta(T_{\mathrm{peak}})=1$, and the sign of the leading nonlinear correction $\Lambda$ changes near this point (more precisely at $\eta_{c}(\sigma)$, as shown by the perturbative analysis in Sec.~\ref{subsec:nlrt_corr} and Figs.~\ref{fig:lambda_xi}, \ref{fig:lambda_eta}), leading to qualitatively different, and even opposite, deviations of LRT on either side of the peak. We emphasize that this sign change has direct practical consequences for MH.
By way of illustration, we use a cobalt nanomagnet as the model system whose large anisotropy constant places $T_{\mathrm{peak}}$ in a cryogenic
but accessible range, thus allowing both sides of the resonance to be explored. However, the formalism presented here is material-independent; in Sec.~\ref{sec:magnetite} we translate the analytical estimates to Fe$_{3}$O$_{4}$ ferrofluids at physiological temperature.

The paper is organised as follows. In Section~\ref{sec:model}, we present the model and the two SAR methods (LLL and LRT), establish their analytical equivalence in the linear regime, and derive the perturbative expression for the leading nonlinear deviation $\Lambda$ together with its physical interpretation.
In Section~\ref{sec:results}, we discuss the numerical results: the Debye resonance and SAR versus temperature, the amplitude and frequency sweeps characterising $\Lambda(\xi,\eta)$ on both sides of the resonance, and an analytical estimate for Fe$_{3}$O$_{4}$ ferrofluids at physiological temperature.
Section~\ref{sec:conclusion} summarises our conclusions.


\section{Model and methods}
\label{sec:model} 

\subsection{Stochastic equation of motion}
\label{subsec:LLE}

We consider a single uniaxial nanomagnet of volume $V$, anisotropy constant $K_{2}$, and saturation moment $\mu_{s}=M_{s}V$. The AC field $\bm{H}_{\mathrm{AC}}(t)=H_{0}\cos(\om t)\,\hat{\bm{e}}_{z}$ is applied in the direction of the easy axis.
The unit magnetisation vector $\bm{m}$ obeys the LLG equation in Landau--Lifshitz form,
\begin{equation}
\frac{d\bm{m}}{d\tau}=-\frac{1}{1+\alpha^{2}}\bigl[\bm{m}\times\bm{h}_{\mathrm{eff}}+\alpha\,\bm{m}\times(\bm{m}\times\bm{h}_{\mathrm{eff}})\bigr],\label{eq:LLGE}
\end{equation}
where $\tau=t/\tsc$ with $\tsc=(\alpha\gamma\mu_0\Ha)^{-1}$ ($\gamma$ is the gyromagnetic ratio in rad\,s$^{-1}$T$^{-1}$), $\Ha=2K_{2}V/(\mu_0\mu_{s})=2K_2/(\mu_0M_s)$ is the anisotropy field, and $\bm{h}_{\mathrm{eff}}=\bm{h}_{\mathrm{an}}+\bm{h}_{\mathrm{AC}}+\bm{h}_{\mathrm{th}}$ is the dimensionless effective field\footnote{Throughout, magnetic fields are in A\,m$^{-1}$; the corresponding flux density (in tesla) is $\mu_0H$, e.g.\ $B_0=\mu_0H_0$ and $\mu_0\Ha=2K_2/M_s$. Zeeman-type energies are thus $\mu_0\mu_sH$, with $\mu_s=M_sV$. In each contribution to the effective field, we use the notation $h\equiv H/H_a$.}. In the LLL approach, thermal fluctuations enter as a Gaussian white-noise field $\bm{h}_{\mathrm{th}}$, the Langevin field, with variance $D=\alpha^{2}/[(1+\alpha^{2})\sig]$ per unit time step~\cite{Brown1963stoch, GarciaPalacios1998}, where $\sig=K_{2}V/k_{B}T$ is the reduced energy barrier. The Landau--Lifshitz equation (without the $1+\alpha^{2}$ gyromagnetic renormalisation) is recovered in the limit $\alpha\to0$~\cite{GarciaPalacios1998, garpal00acp, DejardinKachkachiMartinez2012Comment}. We work at fixed $\alpha=1$ throughout and accordingly use the intermediate-to-high damping approximation for the relaxation rate.
This value places the simulation at the lower boundary of the
intermediate-to-high damping (IHD) regime, ensuring the validity of the relaxation-time formula~\eqref{eq:tauN}, and is consistent with the
landmark stochastic-LLG benchmark of Ref.~\onlinecite{GarciaPalacios1998}.
More importantly, the perturbative correction formula~\eqref{eq:Lambda_3}
derived below is \emph{independent} of $\alpha$: the damping parameter
enters only through $\tau_{\mathrm{N}}(\alpha)$ in the product
$\eta=\omega\tau_{\mathrm{N}}$, and cancels identically in the susceptibility
ratio $\Im[\chi^{(3)}(\omega)]/\Im[\chi^{(1)}(\omega)]$ from which $\Lambda$
is obtained. All conclusions about the sign-change threshold $\eta_{c}(\sigma)$
and the three-parameter map $\Lambda(\sigma,\xi,\eta)$ therefore hold for
any $\alpha$ in the IHD regime; varying $\alpha$ merely re-maps the physical
coordinates $(T,f,V,K_{2})$ onto the universal $(\sigma,\xi,\eta)$ axes.

We introduce the dimensionless parameters
\begin{equation}
\sigma=\frac{K_{2}V}{k_{B}T},\quad h_{0}=\frac{B_{0}}{\mu_{0}\Ha},\quad\xi=\frac{\mu_{s}B_{0}}{k_{B}T}=2\sig h_{0},\label{eq:params}
\end{equation}
and $\eta=\om\ta$ where $\ta$ is the zero-field Néel relaxation time (Sec.~\ref{subsec:LRT}).

\subsection{SAR from the loop-area using LLL}
\label{subsec:LLL}

The energy absorbed per AC cycle equals the hysteresis-loop area.
Projecting onto the field direction, we get
\begin{equation}
\SAR_{\LLL}=\frac{f\,\mu_{0}\mu_{s}\,\Ha\,h_{0}}{V\,\rho}\,\left\langle \om\!\int_{0}^{T_{\mathrm{per}}}m_{z}(\tau)\sin(\om\tau)\,d\tau\right\rangle _{\mathrm{traj}},\label{eq:SAR_LLL}
\end{equation}
where the average is over $N_{\mathrm{traj}}$ independent stochastic trajectories, each integrated for $N_{\mathrm{cyc}}$ stationary cycles after a transient of $N_{\mathrm{tr}}$ cycles (the integral over one period $T_{\mathrm{per}}$ being averaged over these cycles). The signal-to-noise ratio scales as $h_{0}\sqrt{N_{\mathrm{traj}}N_{\mathrm{cyc}}}$, which leads to the constraint on sampling: $N_{\mathrm{traj}}N_{\mathrm{cyc}}\gtrsim(10/h_{0,\min})^{2}$.
The variance is reduced by antithetic-variate sampling~\cite{Hammersley1956, Owen2008} (pairs of $\pm$noise trajectories) and all pair-loops are parallelised with OpenMP. In our runs, $N_{\mathrm{traj}}$ ranges from $\sim10^{2}$ to $5\times10^{3}$, depending on the figure, and $N_{\mathrm{cyc}}\sim50$. Given the scaling above, the relative statistical error on $\SAR_{\LLL}$ decreases as $1/\xi$. As an order of magnitude, estimated from the point-to-point scatter of the data, it is of order $10^{-3}/\xi$ in the fast regime ($T=50$\,K, $N_{\mathrm{traj}}=5\times10^{3}$) and of order $10^{-2}$ in the blocked regime ($T=25$\,K, $N_{\mathrm{traj}}\sim10^{3}$), where slow barrier crossings reduce the effective number of independent samples. These errors are small compared to the deviations $\Lambda$ discussed below, except at the smallest amplitudes, and are sufficient for the qualitative comparison with LRT made here.

The stochastic LLGE~\eqref{eq:LLGE} is integrated numerically using a fourth-order Runge--Kutta (RK4) scheme in which a single realisation of the thermal field $\bm{h}_{\mathrm{th}}$, a Gaussian vector with zero mean and variance $\langle h_{\mathrm{th},i}(\tau)\,h_{\mathrm{th},j}(\tau')\rangle =2D\,\delta_{ij}\delta(\tau-\tau')$, is drawn once per step and held constant through all four sub-stages.
By holding the noise frozen within each step, we make the local increment consistent with the Stratonovich prescription, which is the physically correct interpretation for the magnetic Langevin equation~\cite{garpal00acp}.
The dimensionless time step is set adaptively to $\Delta\tau = T_{\mathrm{per}}/N_{\mathrm{step}}$ with $N_{\mathrm{step}} = 1000$ steps per AC period (capped at $\Delta\tau_{\max}= t_{\mathrm{s}}$ to stay within the stability region of the deterministic LLG). We note that step-size convergence was verified by doubling $N_{\mathrm{step}}$, and we have observed that the SAR changed by less than $0.5\%$ across all field amplitudes and temperatures reported here.

\subsection{SAR from linear response theory (LRT)}
\label{subsec:LRT}

Within LRT~\cite{KuboTodaHashitsume1991, Rosensweig2002, Dejardin2017, DejKac2022},
\begin{equation}
\SAR_{\LRT}=\frac{\mu_{0}\,\om\,\chipp(\om)\,H_{0}^{2}}{2\rho},\qquad\chipp=\chiz\,\frac{\eta}{1+\eta^{2}},\label{eq:SAR_LRT}
\end{equation}
with $\chiz\equiv\mu_{0}\mu_{s}^{2}\langle\cos^{2}\theta\rangle_{\sig}/(k_{B}TV)$ being the static susceptibility, where the Boltzmann average at reduced anisotropy $\sigma$ is given by
\begin{equation}
  \langle\cos^2\theta\rangle_\sigma
    = \frac{\displaystyle\int_{-1}^{1} u^2\, e^{\sigma u^2}\,du}
           {\displaystyle\int_{-1}^{1} e^{\sigma u^2}\,du}
\label{eq:boltz}
\end{equation}
with limits: $\sigma\to 0 \Rightarrow \langle\cos^2\theta\rangle\to\tfrac{1}{3}$ (isotropic
Langevin); $\sigma\gg 1 \Rightarrow \langle\cos^2\theta\rangle\to 1-1/\sigma$ (blocked).

Regarding the relaxation time appearing in $\eta=\omega\ta$, in practice, the Aharoni-Néel-Brown expression is only valid for $h\leq0.4$ and $\sigma\geq2$ {[}see Ref. \onlinecite{CoffeyEtal_PhysRevB.51.15947}{]}. Cregg, Crothers and Wickstead~\citep{creetal94jap} proposed an improved expression that is valid in the whole range of $\sigma$:
\begin{equation}
\ta = (1+\alpha^2)\tau_{s}\left(\frac{2}{\sqrt{\pi}}\frac{\sigma^{3/2}}{1+\sigma} + 2^{-\sigma}\right)^{-1}\left(e^{\sigma} -1\right).
 \label{eq:tauN}
\end{equation}

%
As mentioned earlier, the prefactor $(1+\alpha^{2})$ reflects the choice of dynamics (LLGE); for $\alpha\to0$ (Landau--Lifshitz limit) it reduces to unity and $\tsc\to(\gamma\mu_0\Ha)^{-1}$.

%

\subsection{Consistency of LRT and LLL}
\label{subsec:consistency}
The magnetization of a system in the direction of the probing magnetic field $H$ can be written as
\[
m=\chi_{1}H+\chi_{3}H^{3}+\chi_{5}H^{5}+\ldots=\sum_{n\geq0}\chi_{2n+1}H^{2n+1}\left(t\right).
\]
Then, under the excitation $\mathbf{H}=h_0\,e^{i\omega t}\mathbf{e}_{h}$, we obtain the nonlinear magnetic response ($m_{n}=m_{n}^{\prime}-im_{n}^{\prime\prime}$)\cite{BitohEtal_JPSJ.64.1311, BitohEtal_J3M.154.59}:
\begin{align}
&m\left(t\right)  =\Re\left[\sum_{n\geq0}m_{2n+1}e^{i\left(2n+1\right)\omega t}\right]\label{eq:MagExp2}\\
& =\sum_{n\geq0}\Bigg[
m_{2n+1}^{\prime}\cos\left[\left(2n+1\right)\omega t\right]
+m_{2n+1}^{\prime\prime}\sin\left[\left(2n+1\right)\omega t\right]
\Bigg]
\nonumber
\end{align}
where
\begin{align*}
m_{1}^{\prime} & =\chi_{1}^{\prime}h_0+\frac{3}{4}\chi_{3}^{\prime}h_0^{3}+\ldots, m_{1}^{\prime\prime}=\chi_{1}^{\prime\prime}h_0+\frac{3}{4}\chi_{3}^{\prime\prime}h_0^{3}+\ldots, \\
m_{3}^{\prime} & =\frac{1}{4}\chi_{3}^{\prime}h_0^{3}+\frac{5}{16}\chi_{5}^{\prime}h_0^{5}+\ldots, m_{3}^{\prime\prime}=\frac{1}{4}\chi_{3}^{\prime\prime}h_0^{3}+\frac{5}{16}\chi_{5}^{\prime\prime}h_0^{5}+\ldots.
\end{align*}
Here $\chi_{1}(\om)$ and $\chi_{3}(\om)\equiv\chi_{3}(\om;\om,\om,-\om)$ are the linear and cubic susceptibilities at the fundamental frequency; the weight $3/4$ is that of the fundamental in $\cos^{3}\omega t=(3\cos\omega t+\cos3\omega t)/4$.

Next, since $\int_{0}^{T_{\mathrm{per}}}\cos(n\om t)\sin(\om t)\,dt=0$ and $\int_{0}^{T_{\mathrm{per}}}\sin(n\om t)\sin(\om t)\,dt=(\pi/\om)\,\delta_{n1}$, only the out-of-phase fundamental component $m_{1}^{\prime\prime}$ of Eq.~\eqref{eq:MagExp2} contributes to the loop area: the in-phase components and all higher harmonics enclose no area. The sine integral in Eq.~(\ref{eq:SAR_LLL}) is thus $I=\pi m_{1}^{\prime\prime}/\om$ and $\SAR_{\LLL}\propto m_{1}^{\prime\prime}$. In the limit $h_{0}\to0$, $m_{1}^{\prime\prime}\to h_{0}\chipp_1$ and $I=h_{0}\chipp_1\pi/\om$, where $\chipp_1$ is the (linear) out-of-phase susceptibility. Converting to SI via $h_{0}=H_{0}/\Ha$ and $\chipp_{\mathrm{SI}}=M_{s}\,\chipp_1/\Ha$ one recovers Eq.~(\ref{eq:SAR_LRT}) exactly~\cite{Coffey1998}.

This proves that $\SAR_{\LLL}$ and $\SAR_{\LRT}$ are consistent methods: any discrepancy at finite $h_{0}$, for instance, marks a nonlinear effect.

\subsection{Deviation of LLL from LRT: perturbative expansion}
\label{subsec:nlrt_corr}
%
To quantify the comparison between the two approaches, we introduce the relative deviation
\begin{equation}
\Lambda(\xi)\equiv\frac{\SAR_{\LLL}}{\SAR_{\LRT}}-1.\label{eq:Lambda}
\end{equation}

In the next section, we discuss the behavior of $\Lambda$ as obtained by the LLL approach and compare it to the approximate analytical expression we derive below from LRT.

Since only $m_{1}^{\prime\prime}$ contributes to $\SAR_{\LLL}$, Eq.~\eqref{eq:MagExp2} gives, to first order beyond LRT, $\SAR_{\LLL}\propto\chi_{1}^{\prime\prime}h_{0}+\tfrac{3}{4}\chi_{3}^{\prime\prime}h_{0}^{3}$. Writing the response in terms of $\xi$, $m=\chi^{(1)}\xi+\chi^{(3)}\xi^{3}+\ldots$ (so that $\chi_{k}h_{0}^{k}=\chi^{(k)}\xi^{k}$), and using $\chi''=-\Im\chi$, we obtain
\begin{equation}
\Lambda\simeq\frac{3}{4}\,\xi^{2}\,\frac{\Im\chi^{(3)}(\om)}{\Im\chi^{(1)}(\om)},
\label{eq:Lambda_2}
\end{equation}
where $\chi^{(3)}(\om)$ is the cubic susceptibility at the fundamental frequency. Both susceptibilities follow from solving, order by order in the field, the relaxation equation of the longitudinal magnetisation of a uniaxial nanomagnet (two-well, high-barrier regime) with the field-dependent rate $\Gamma=\Gamma_{0}(1+\xi^{2}/2)$, $\Gamma_{0}=1/\ta$~\cite{garpalgar04prb, JonssonGarciaPalacios2001, vernayetal14prb}. With $\eta=\om\ta$, this yields
\begin{align}
\chi^{(1)}(\om)&=\frac{\chi_{\mathrm{eq}}^{(1)}}{1+i\eta},\nonumber\\
\chi^{(3)}(\om)&=\frac{1}{1+i\eta}\left[\chi_{\mathrm{eq}}^{(3)}+\frac{\chi_{\mathrm{eq}}^{(1)}}{6}\,\frac{3\eta^{2}+i\eta}{1+\eta^{2}}\right],
\label{eq:chi13}
\end{align}
where the second term in the bracket comes from the field dependence of $\Gamma$.
The static equilibrium susceptibilities, $\chi_{\mathrm{eq}}^{(1)}$ and $\chi_{\mathrm{eq}}^{(3)}$, for the longitudinal geometry, are given exactly by the Boltzmann averages of Eq.~(\ref{eq:boltz}), $\chi_{\mathrm{eq}}^{(1)}=\langle\cos^2\theta\rangle_\sigma$ and $\chi_{\mathrm{eq}}^{(3)}=[\langle\cos^4\theta\rangle_\sigma-3\langle\cos^2\theta\rangle_\sigma^2]/6$. Here we use their high-barrier (large-$\sigma$) forms, which can be obtained from Eq.~(8) of Ref.~\onlinecite{vernayetal14prb} (upon setting there $\zeta=0$, $x=0$) by retaining the leading terms in $1/\sigma$:
\begin{equation}
\chi_{\mathrm{eq}}^{(1)}=1-\frac{1}{\sigma},\qquad
\chi_{\mathrm{eq}}^{(3)}=\frac{1}{3}\!\left(-1+\frac{2}{\sigma}\right).
\label{eq:chieq}
\end{equation}
Inserting Eqs.~\eqref{eq:chi13} and \eqref{eq:chieq} into Eq.~\eqref{eq:Lambda_2} yields
\begin{equation}
\Lambda(\sigma, \xi, \eta) \simeq -\xi^2\,F(\sigma, \eta),\quad
F(\sigma, \eta) = \frac{\sigma-2}{4(\sigma-1)}+\frac{1-3\eta^{2}}{8(1+\eta^{2})}.
\label{eq:Lambda_3}
\end{equation}

Equation~(\ref{eq:chieq}) is thus an approximation, valid for $\sigma\gg1$ (these expressions were studied in Ref.~\onlinecite{vernayetal14prb}, to which we refer for details). The ratio $\chi_{\mathrm{eq}}^{(3)}/\chi_{\mathrm{eq}}^{(1)}$, which is what enters $\Lambda$, differs from its exact value by about $6\%$ at $\sigma=3.2$, $2.4\%$ at $\sigma=5.3$, and less than $1\%$ for $\sigma\gtrsim11$, so that Eq.~(\ref{eq:Lambda_3}) is reliable for $\sigma\gtrsim3$, which covers all the cases considered here. For $\sigma\lesssim2$ the approximation breaks down ($\chi_{\mathrm{eq}}^{(1)}$ vanishes at $\sigma=1$ and $\chi_{\mathrm{eq}}^{(3)}$ changes sign at $\sigma=2$, whereas the exact expressions remain finite and $\chi_{\mathrm{eq}}^{(3)}<0$ for all $\sigma$); there, the exact averages should be used, \textit{i.e.}, $F=-\tfrac{3}{4}\,\chi_{\mathrm{eq}}^{(3)}/\chi_{\mathrm{eq}}^{(1)}+\tfrac{1}{8}(1-3\eta^2)/(1+\eta^2)$ in Eq.~(\ref{eq:Lambda_3}).

In the two limiting regimes this gives
\begin{equation}
\Lambda \simeq
\begin{cases}
-\dfrac{3\sigma-5}{8(\sigma-1)}\,\xi^{2}, & \eta \ll 1 \quad \text{(fast regime)}, \\[1.2em]
+\dfrac{\sigma+1}{8(\sigma-1)}\,\xi^{2}, & \eta \gg 1 \quad \text{(blocked regime)}.
\end{cases}
\label{eq:lam_cubic}
\end{equation}
so that $\Lambda<0$ (LLL below LRT) in the fast regime and $\Lambda>0$ (LLL above LRT) in the blocked regime. $\Lambda$ changes sign at $\eta_{c}=\sqrt{(3\sigma-5)/(\sigma+1)}$, which tends to $\sqrt{3}$ for $\sigma\to\infty$ and equals $1.05$, $1.33$ and $1.53$ for $\sigma=3.2$, $5.5$ and $11.1$, respectively.

We note in passing that if the field dependence of the relaxation rate were neglected ($\Gamma\longrightarrow \Gamma_0$), Eq.~\eqref{eq:chi13} would give $\Lambda=-\xi^{2}(\sigma-2)/[4(\sigma-1)]$ at all $\eta$, i.e., no sign change. Therefore, the positive $\Lambda$ in the blocked regime is a direct consequence of the field-induced increase of $\Gamma$.


The Debye susceptibility $\chipp(\om)=\chiz\,\eta/(1+\eta^{2})$ is maximised at $\eta=\om\ta=1$ and decreases on both sides. Then, as is well known, an applied field modifies the \emph{effective} barrier height and thereby the \emph{effective} relaxation time $\ta^{\mathrm{eff}}(h_{0})$.
For a uniaxial nanomagnet with the field along the easy axis, the barrier is lowered by the Zeeman energy: $\Delta E_{\pm}\propto\sig(1\mp h_{0})^{2}$. Therefore $\ta^{\mathrm{eff}}(h_{0})<\ta(0)$, \textit{i.e.}, the field always moves $\omega\tau_{\mathrm{N}}$ downward relative to its zero-field value. Consequently, we observe the following behavior [illustrated in Figure~\ref{fig:sar_T}],
\begin{itemize}
\item If $\om\ta(0)<1$ (fast regime, $T>T_{\mathrm{peak}}$), moving further below resonance reduces $\chipp_{\mathrm{eff}}$ relative to
the LRT prediction, leading to $\Lambda<0$.
\item If $\om\ta(0)>1$ (blocked regime, $T<T_{\mathrm{peak}}$): the field
drives $\omega\tau_{\mathrm{eff}}$ toward resonance, increasing $\chipp_{\mathrm{eff}}$
above the LRT prediction and thereby $\Lambda>0$.
\end{itemize}

This effective-$\tau$ picture locates the sign change of $\Lambda$ at $\eta=1$. The perturbative formula in Eq.~\eqref{eq:Lambda_3} refines this: the threshold is $\eta_{c}(\sigma)$ given above, see Fig.~\ref{fig:lambda_eta}.

\section{Results}
\label{sec:results}
For a quantitative discussion of the results, we use the set of physical parameters associated with a cobalt nanomagnet\cite{jametetal01prl, jametetal04prb} with diameter $d=3\,\mathrm{nm}$: $K_{2}=2.7\times10^{5}\,\mathrm{J\,m^{-3}}$, $M_{s}=1.68\times10^{6}\,\mathrm{A\,m^{-1}}$, anisotropy field $\mu_{0}\Ha=2K_{2}/M_{s}=0.322\,\mathrm{T}$, and AC frequency $f=10\,\mathrm{MHz}$; this yields $\tsc=1.76\times10^{-11}\,\mathrm{s}$.
Then, the resonance condition $\om\ta(T_{\mathrm{peak}})=1$ is satisfied at $T_{\mathrm{peak}}\simeq\SI{39}{\kelvin}$.
Cobalt is used here as a theoretical model system, not as a clinical candidate. Its large magnetocrystalline anisotropy constant places $T_{\mathrm{peak}}$ in a cryogenic but experimentally accessible range~\cite{jametetal01prl, jametetal04prb}, and this is precisely the feature that allows both dynamical regimes---fast-relaxation ($T>T_{\mathrm{peak}}$) and blocked ($T<T_{\mathrm{peak}}$), to be explored systematically within a single material system. The formalism is material-independent; the clinically relevant case of Fe$_{3}$O$_{4}$ nanoparticles at physiological temperature $T=310\,\mathrm{K}$ is addressed in Sec.~\ref{sec:magnetite}.
Above $T_{\mathrm{peak}}$ the system is in the \emph{fast-relaxation}
regime ($\om\ta<1$), \textit{i.e.} the superparamagnetic regime which is relevant to MH applications; below it lies in the \emph{blocked} regime ($\om\ta>1$).

\subsection{SAR versus temperature}
\label{sec:sar_T}

\begin{figure}[ht!]
\centering \includegraphics[width=1\columnwidth]{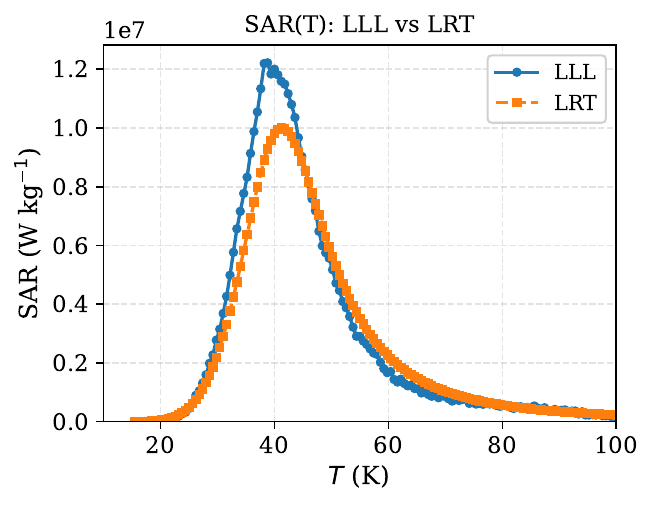}
\caption{SAR versus temperature at $B_{0}=10\,\mathrm{mT}$, $f=10\,\mathrm{MHz}$.
Circles (solid line): LLL stochastic simulation. Squares (dashed line): LRT with the IHD Néel time.}
\label{fig:sar_T} 
\end{figure}

Figure~\ref{fig:sar_T} shows $\SAR_{\LLL}(T)$ and $\SAR_{\LRT}(T)$ at fixed amplitude $B_{0}=10\,\mathrm{mT}$ ($h_{0}\simeq0.031$,
$\xi\simeq0.44$ at $T_{\mathrm{peak}}$).
We see that both approaches produce a Debye peak: LLL at $T_\mathrm{peak}^\mathrm{LLL}\simeq39\,\mathrm{K}$, in good agreement with the CCW prediction in Eq.~(\ref{eq:tauN}), and LRT (evaluated with the IHD formula) at $T_\mathrm{peak}^\mathrm{LRT}\simeq41\,\mathrm{K}$, shifted upward by the mild IHD overestimate of $\tau_N$. The peak height differs by $\sim20\%$: LLL exceeds LRT in the blocked regime ($T<T_\mathrm{peak}, \Lambda>0)$, which is consistent with Eq.~(\ref{eq:lam_cubic}) at $\xi(T)\simeq0.44-1.1$, while LRT overestimates the SAR above $T_\mathrm{peak}$, where the fast-relaxation mechanism reverses the sign of $\Lambda$ (Sec.~\ref{subsec:nlrt_corr}).
At temperatures far from the peak on either side, both SAR values decay toward zero and the absolute difference becomes indistinguishable on a linear scale, though the relative deviation remains large.

\subsection{LRT validity as a function of AC amplitude and frequency}
\label{subsec:sar_h0}
\begin{figure*}[ht!]
\centering \includegraphics[width=\textwidth]{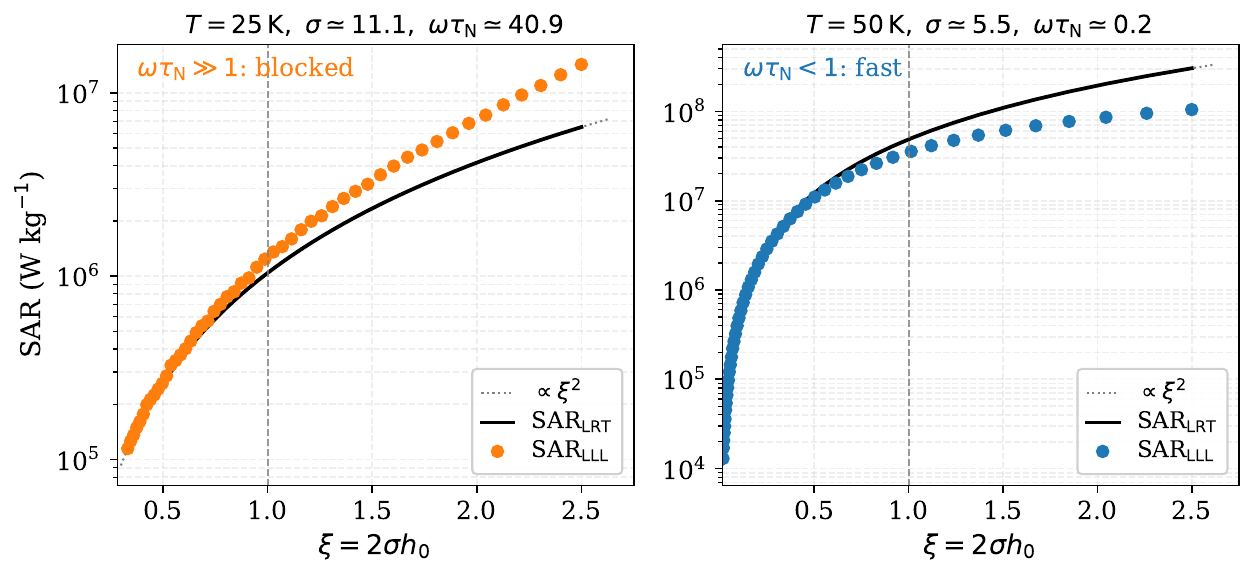}
\caption{$\SAR_{\LLL}$ (symbols) and $\SAR_{\LRT}$ (solid line) versus $\xi$
on linear--log axes. Left: $T=25\,\mathrm{K}$ ($\om\ta\simeq41$,
blocked regime), LLL exceeds LRT for $\xi\gtrsim0.3$. Right:
$T=50\,\mathrm{K}$ ($\om\ta\simeq0.25$, fast regime), LLL falls
below LRT for $\xi\gtrsim0.3$. Dotted: $\xi^{2}$ reference. Dashed
vertical: $\xi=1$ (LRT threshold $h_{0}^{*}=1/(2\sig)$).}
\label{fig:sar_xi}
\end{figure*}
We have varied the AMF amplitude at two temperatures chosen on opposite sides of $T_{\mathrm{peak}}$: $T=50\,\mathrm{K}$ (fast-relaxation,
$\sig\simeq5.5$, $\om\ta\simeq0.25$) and $T=25\,\mathrm{K}$ (blocked, $\sig\simeq11.1$, $\om\ta\simeq41$).
Figure~\ref{fig:sar_xi} shows $\SAR_{\LLL}$ and $\SAR_{\LRT}$ as functions of $\xi=2\sig h_{0}$ on linear--log axes. Both panels display the expected quadratic scaling $\SAR\propto\xi^{2}$ at small $\xi$, confirming the linear regime. More precisely, the two methods agree near $\xi=0$ and progressively diverge as $\xi$ grows, but the direction of the divergence is opposite for the two temperatures: at $T=50\,\mathrm{K}$ the LLL curve falls below LRT, while at $T=25\,\mathrm{K}$ it rises above.

\begin{figure}[ht!]
\centering \includegraphics[width=1\columnwidth]{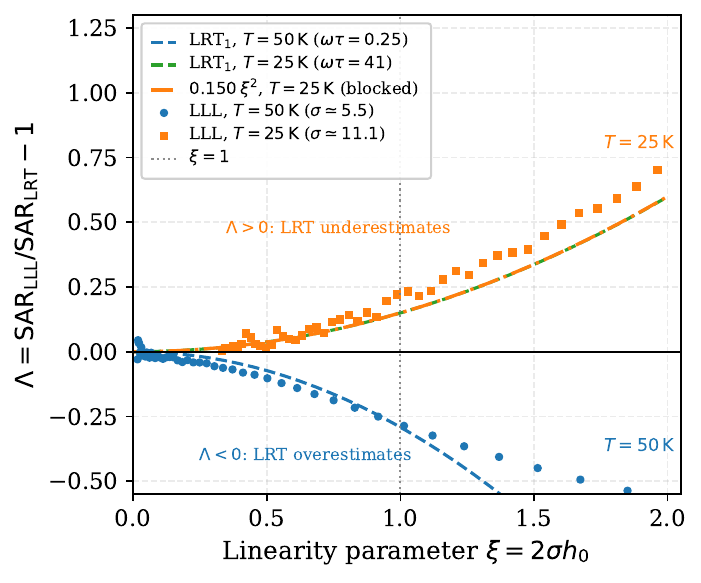}
\caption{Relative deviation $\Lambda=\SAR_{\LLL}/\SAR_{\LRT}-1$ versus $\xi$.
Symbols: LLL simulation (circles: $T=50\,\mathrm{K}$, $\sigma\simeq5.5$;
squares: $T=25\,\mathrm{K}$, $\sigma\simeq11.1$). Dashed lines:
first-order perturbative prediction Eq.~(\ref{eq:Lambda_3}). Solid
(orange): blocked-regime analytical formula Eq.~(\ref{eq:lam_cubic}),
valid for $\om\ta\gg1$
($T=25\,\mathrm{K}$, $\sigma\simeq11.1$, prefactor $\simeq0.150$).
Dotted vertical: $\xi=1$.
}
\label{fig:lambda_xi}
\end{figure}

Figure~\ref{fig:lambda_xi} shows a comparison between LLL and the perturbative approximation in Eq.~(\ref{eq:Lambda_3}).
The perturbative prediction reproduces the sign and the order of magnitude of the LLL data. In the fast regime ($T=50\,\mathrm{K}$, $\om\ta\simeq0.25$), it underestimates $|\Lambda|$ for $\xi\lesssim0.7$ (by a factor $\simeq1.4$ at $\xi=0.5$), agrees with the LLL data to within $5\%$ at $\xi\simeq1$, and overestimates $|\Lambda|$ beyond (by a factor $\simeq2$ at $\xi=2$), since the cubic prediction $-\xi^{2}F$ grows without bound while the true deviation is bounded ($\Lambda\geq-1$, because $\SAR_{\LLL}\geq0$).
In the blocked regime ($T=25\,\mathrm{K}$, $\om\ta\simeq41$), the large-$\eta$ limit Eq.~(\ref{eq:lam_cubic}) (solid orange line, prefactor $\simeq0.150$) lies below the LLL data, while for smaller $\xi$ both deviations are below $\sim0.04$, which is comparable to the statistical error in this regime. A quantitative agreement is not to be expected, given the approximations involved in Eq.~(\ref{eq:Lambda_3}) (high-barrier forms, truncation at order $\xi^{2}$) and the statistical and numerical errors of the simulation.
In both cases, the LLL simulation provides the more reliable quantitative result beyond the strict linear regime.
The physical mechanism underlying the sign reversal and its location at $\eta_{c}(\sigma)$ were analysed in Sec.~\ref{subsec:nlrt_corr}.

\begin{figure}[ht!]
\centering\includegraphics[width=1\columnwidth]{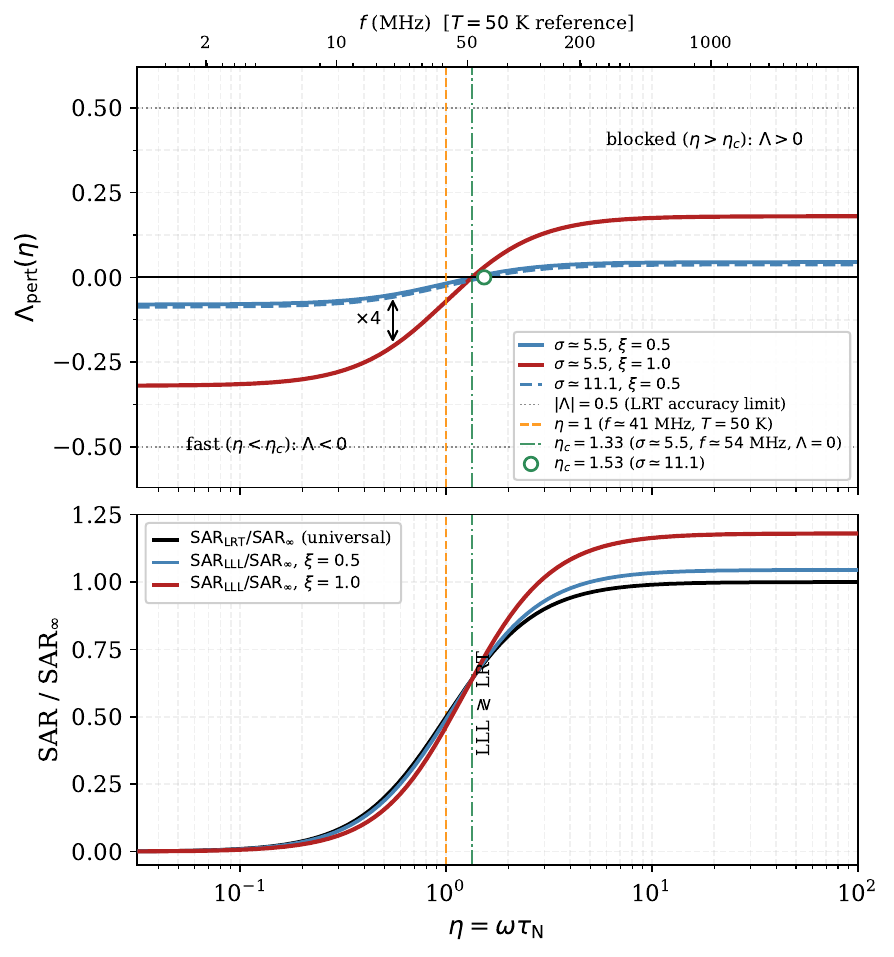}
\caption{Top: perturbative deviation $\Lambda_\mathrm{pert}(\eta)$ from Eq.~\eqref{eq:Lambda_3} versus $\eta=\omega\tau_\mathrm{N}$ at fixed $\sigma$ and $\xi$. Solid curves: $\sigma\simeq5.5$ ($T=50\,\mathrm{K}$) for $\xi=0.5$ (blue) and $\xi=1.0$ (red). Dashed: $\sigma\simeq11.1$ ($T=25\,\mathrm{K}$), $\xi=0.5$. All curves change sign at $\eta_{c}(\sigma)$, independent of $\xi$: $\eta_{c}=1.33$ for $\sigma\simeq5.5$ (dash-dot green vertical) and $\eta_{c}=1.53$ for $\sigma\simeq11.1$ (open green circle). Orange dashed vertical: $\eta=1$ (Debye resonance). Dotted horizontal: $|\Lambda|=0.5$ (accuracy limit of LRT). The secondary $x$-axis gives the frequency $f$ for the $T=50\,\mathrm{K}$ reference ($\tau_\mathrm{N}\simeq3.9\,\mathrm{ns}$).
Bottom: normalised $\mathrm{SAR}/\mathrm{SAR}_\infty$ versus $\eta$. Black: universal LRT curve $\eta^2/(1+\eta^2)$. Blue/red: LLL perturbative estimate for $\xi=0.5$ and $\xi=1.0$ ($\sigma\simeq5.5$). Both LLL curves cross the LRT curve at $\eta_{c}(\sigma\simeq5.5)=1.33$.}
\label{fig:lambda_eta}
\end{figure}
Figure~\ref{fig:lambda_eta} illustrates the full $\eta$-dependence of Eq.~\eqref{eq:Lambda_3} at fixed $\sigma$ and $\xi$.
The top panel shows $\Lambda_\mathrm{pert}(\eta)$ for the Co nanomagnet parameters at $T=50\,\mathrm{K}$ ($\sigma\simeq5.5$, $\tau_\mathrm{N}\simeq3.9\,\mathrm{ns}$), for two field amplitudes $\xi=0.5$ and $\xi=1.0$, and also for $T=25\,\mathrm{K}$ ($\sigma\simeq11.1$) at $\xi=0.5$.
Three features are immediately visible.
First, $\Lambda$ changes sign at $\eta_{c}(\sigma)=\sqrt{(3\sigma-5)/(\sigma+1)}$, a property of $F(\sigma,\eta)$ alone, independent of $\xi$ and weakly dependent on $\sigma$: $\eta_{c}=1.33$ for $\sigma\simeq5.5$ and $1.53$ for $\sigma\simeq11.1$, and $\eta_{c}\to\sqrt{3}$ as $\sigma\to\infty$.
Second, the magnitude scales strictly as $\xi^2$: the $\xi=1.0$ curve is everywhere exactly four times larger in $|\Lambda|$ than the $\xi=0.5$ curve; this ratio is marked by the double arrow in the figure.
Third, the asymptotic values recover Eq.~\eqref{eq:lam_cubic}: the fast-regime plateau ($\eta\ll1$) is $-\frac{3\sigma-5}{8(\sigma-1)}\xi^2$, and the blocked plateau ($\eta\gg1$) is $+\frac{\sigma+1}{8(\sigma-1)}\xi^2$.
Note that the sign change of $\Lambda$ occurs at $\eta_{c}>1$, which is distinct from the Debye resonance at $\eta=1$ where $\chi''(\omega)$ peaks.
For the Co system at $T=50\,\mathrm{K}$ these correspond to $f\simeq41\,\mathrm{MHz}$ ($\eta=1$) and $f\simeq54\,\mathrm{MHz}$ ($\eta=\eta_{c}=1.33$), respectively.
The bottom panel shows the normalised SAR, $\mathrm{SAR}/\mathrm{SAR}_\infty$ where $\mathrm{SAR}_\infty \equiv \lim_{\eta\to\infty}\mathrm{SAR}_\mathrm{LRT}$, as a function of $\eta$.
The LRT result (black curve, $\eta^2/(1+\eta^2)$) is universal and independent of $\xi$.
The LLL curves cross the LRT curve at $\eta_{c}$: below this value LLL lies beneath LRT ($\Lambda<0$, LRT overestimates); above it LLL exceeds LRT ($\Lambda>0$, LRT underestimates).

\subsection{\texorpdfstring{Estimates for Fe$_{3}$O$_{4}$ ferrofluids}{Estimates for Fe3O4 ferrofluids}}
\label{sec:magnetite}
\begin{figure}[ht!]
\centering \includegraphics[width=1\columnwidth]{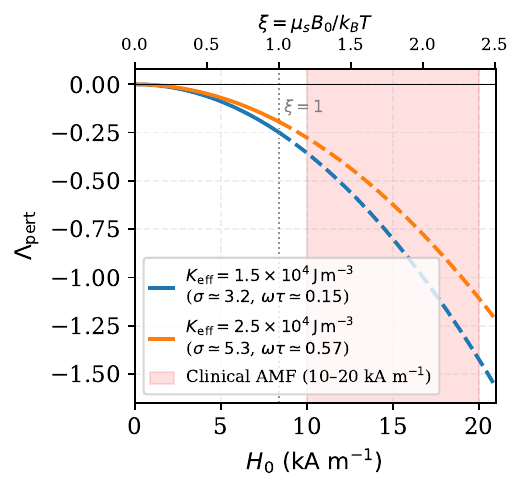}
\caption{Perturbative estimate $\Lambda_{\mathrm{pert}}$ (Eq.~\ref{eq:Lambda_3})
versus AC amplitude $H_{0}$ for Fe$_{3}$O$_{4}$ nanomagnets ($d=12\,\mathrm{nm}$,
$T=310\,\mathrm{K}$, $f=10\,\mathrm{MHz}$, $\alpha=1$). Two values
of $K_{\mathrm{eff}}$ bracket the experimental uncertainty. Solid:
valid perturbative range ($\xi<1$, i.e., $H_{0}<8.4\,\mathrm{kA\,m^{-1}}$);
dashed: extrapolation (overestimates $|\Lambda|$). Red band: typical
clinical AMF range. Upper axis: $\xi=\mu_{s}B_{0}/k_{B}T$ (material-independent
for fixed $M_{s}$, $V$, $T$).}
\label{fig:lambda_mag}
\end{figure}

Magnetite (Fe$_{3}$O$_{4}$) and maghemite ($\gamma$-Fe$_{2}$O$_{3}$) are the prototypical materials for MH~\cite{Dutz2014,Fortin2007,Serantes2018}. By way of illustration, let us calculate the perturbative estimate in Eq.~(\ref{eq:Lambda_3}) for a representative 12\,nm Fe$_{3}$O$_{4}$ nanomagnet at body temperature $T=310\,\mathrm{K}$. This diameter is representative of single-domain magnetite nanocrystals produced by standard co-precipitation and thermal decomposition routes, for which the spread in $K_{\mathrm{eff}}$ is captured by the bracketing values used below. Moreover, within linear response theory the SAR is maximised when $\omega\tau_{\mathrm{N}}=1$; since $\tau_{\mathrm{N}}$ grows exponentially with volume, this resonance condition selects an optimal core diameter that, for magnetite under standard clinical AMF conditions, falls at approximately 12\,nm, as demonstrated theoretically by Purushotham and Ramanujan~\cite{PurushothamRamanujan2010} and confirmed experimentally by Narayanaswamy \emph{et al.}~\cite{Narayanaswamy2021}.
The material parameters are $M_{s}=4.5\times10^{5}\,\mathrm{A\,m^{-1}}$, $\rho=5180\,\mathrm{kg\,m^{-3}}$, and $\alpha=1$ (the CCW formula is an IHD approximation, valid only for $\alpha$ not small) and the effective uniaxial anisotropy constant is taken in the range $K_{\mathrm{eff}}=1.5$--$2.5\times10^{4}\,\mathrm{J\,m^{-3}}$, spanning the spread reported for 12\,nm nanocrystals~\cite{Fortin2007,Guardia2012,DeLaPresa2012}.
At $f=10\,\mathrm{MHz}$, $\sig\simeq3.2$--$5.3$ and $\om\ta\simeq0.15$ and $0.57$ for $K_{\mathrm{eff}}=1.5$ and $2.5\times10^{4}\,\mathrm{J\,m^{-3}}$, respectively, placing both systems in the \emph{fast-relaxation} regime ($\om\ta<1$) at body temperature, with a Debye resonance at $T_{\mathrm{peak}}\simeq184\,\mathrm{K}$ and $277\,\mathrm{K}$, both well below $310\,\mathrm{K}$ ($T > T_{\mathrm{peak}}$).
The frequency $f=10\,\mathrm{MHz}$ is not representative of standard clinical protocols, which operate at $f=100\,\mathrm{kHz}$--500\,kHz; it is chosen to place $\eta$ in the range $0.15$--$0.57$, giving an illustration that sits as close as possible to the sign-change threshold $\eta_{c}\simeq1.0$--$1.3$ while remaining in the fast-relaxation regime.
The crucial point is that all results are expressed as functions of the dimensionless ratio $\eta=\om\ta$, which absorbs the frequency dependence entirely.
At clinical frequencies, $\eta$ scales proportionally: $\eta_{\mathrm{clin}}=(f_{\mathrm{clin}}/f)\,\eta_{10\,\mathrm{MHz}}\simeq0.001$--$0.06$, driving the 12\,nm particle deeper into the fast-relaxation plateau of Fig.~\ref{fig:lambda_eta} where $|F(\sigma,\eta)|$ is maximal.
Within Eq.~(\ref{eq:Lambda_3}), the 10\,MHz estimate is therefore a lower bound on $|\Lambda_{\mathrm{pert}}|$ at the same $\xi$ and $\sigma$. Since this expression is quantitatively reliable only for $\xi\lesssim1$ and overestimates $|\Lambda|$ beyond (Fig.~\ref{fig:lambda_xi}), this should be read as an indication that the LRT overestimate at standard clinical conditions is at least comparable to, and probably larger than, that found at 10\,MHz, rather than as a strict bound.

We first note that from the definition $\xi=\mu_{s}B_{0}/k_{B}T$ with $\mu_{s}=M_{s}V$ and $B_0=\mu_0H_0$, the LRT threshold $\xi=1$ corresponds to a field amplitude
\begin{equation}
H_{0}^{*}=\frac{k_{B}T}{\mu_{0}M_{s}V}\simeq8.4\,\mathrm{kA\,m^{-1}},
\end{equation}
which is independent of $K_{\mathrm{eff}}$.
Then, the \textit{in vivo} safety constraint, known as the \emph{Brezovich criterion}~\cite{Brezovich1988,Fraile2024,Dejardin2020},
$H_{0}f\leq4.85\times10^{8}\,\mathrm{A\,m^{-1}\,s^{-1}}$, limits the field to $H_{0}\leq4.85\,\mathrm{kA\,m^{-1}}$ at $f=100\,\mathrm{kHz}$ ($\xi\simeq0.58$). At this amplitude $|\Lambda_\mathrm{pert}|\simeq0.07$ for $K_{\mathrm{eff}}=2.5\times10^4\,\mathrm{J\,m^{-3}}$ and $\simeq0.08$ for $K_{\mathrm{eff}}=1.5\times10^4\,\mathrm{J\,m^{-3}}$, both within the perturbative range.
However, nanomagnets of $d=12\,\mathrm{nm}$ are superparamagnetic at 310\,K and produce negligible SAR at such low amplitudes~\cite{Fraile2024, Kachkachi2026LocalGlobalHeating}. In practice, most experimental and pre-clinical MH protocols exceed the Brezovich limit, operating at $H_{0}=10$--$20\,\mathrm{kA\,m^{-1}}$ [see Refs.~\onlinecite{Dutz2014,Fraile2024}]. In this experimentally realistic regime, $\xi\simeq1.2$--$2.4$, where the perturbative estimate gives $\Lambda_\mathrm{pert}\simeq-0.28$ to $-0.36$ at $\xi=1.2$ and $-0.44$ to $-0.56$ at $\xi=1.5$, and is no longer reliable beyond.

In Figure~\ref{fig:lambda_mag} we plot $\Lambda_{\mathrm{pert}}$ using Eq.~(\ref{eq:Lambda_3}) versus $H_{0}$ for the two bracketing
values of $K_{\mathrm{eff}}$.
At $H_{0}=H_{0}^{*}/2\simeq4.2\,\mathrm{kA\,m^{-1}}$ ($\xi=0.5$), $\Lambda_\mathrm{pert}\simeq-0.06$ and $-0.05$ for $K_{\mathrm{eff}}=1.5$ and $2.5\times10^{4}\,\mathrm{J\,m^{-3}}$, respectively; at $H_{0}=H_{0}^{*}$ ($\xi=1$) they are $-0.25$ and $-0.19$, and $|\Lambda_\mathrm{pert}|$ reaches $0.5$ near $\xi\simeq1.4$ and $1.6$, respectively.
Beyond $\xi=1$ the perturbative formula overestimates $|\Lambda|$ (cf.\ Fig.~\ref{fig:lambda_xi}, $T=50\,\mathrm{K}$ case): the cubic form $-\xi^{2}F$ grows without bound, whereas the true deviation is bounded by $\Lambda\geq-1$ (since $\SAR_{\LLL}\geq0$). The dashed portions of Fig.~\ref{fig:lambda_mag} therefore probably overestimate the true LRT error; LLL or TQMC simulation is needed for quantitative predictions at clinical amplitudes. At lower clinical frequencies ($f=100$\,kHz--1\,MHz), $\om\ta\to0$ and $F$ saturates at its fast-regime plateau value $\frac{3\sigma-5}{8(\sigma-1)}$ (left asymptote in Fig.~\ref{fig:lambda_eta}), so $|\Lambda_{\mathrm{pert}}|$ grows further.

In summary, using LRT to predict SAR for 12\,nm Fe$_{3}$O$_{4}$ at standard MH conditions leads to a
systematic overestimate of heating, by about $5$--$6\%$ at $\xi=0.5$ and $20$--$25\%$ at $\xi=1$ according to the perturbative estimate, which is equivalent to under-dosing when designing treatment protocols.

It is instructive to contrast this with a larger particle. For $d=25\,\mathrm{nm}$, the volume increases by a factor $(25/12)^{3}\simeq9$, so $\sigma(310\,\mathrm{K})\simeq29$--$47$ for the same $K_{\mathrm{eff}}$ range. Inserting these values into the CCW formula~\eqref{eq:tauN} yields $\omega\tau_{\mathrm{N}}\gg1$ at $f=10\,\mathrm{MHz}$ and $T=310\,\mathrm{K}$. Therefore, the 25\,nm particle sits firmly in the \emph{blocked} regime. Consequently, from Eq.~\eqref{eq:lam_cubic}, $\Lambda>0$ and LRT \emph{underestimates} the SAR, and the sign of the LRT error is thus reversed relative to the 12\,nm case. Furthermore, since $\xi\propto V$, the same clinical field amplitude gives $\xi(25\,\mathrm{nm})\simeq9\,\xi(12\,\mathrm{nm})$, placing the system well outside the perturbative range even at modest amplitudes; full LLL or TQMC simulations are then required for quantitative predictions.

\section{Conclusion}

\label{sec:conclusion} 

We have computed and compared the SAR of a uniaxial single-domain nanomagnet as rendered by linear response theory (LRT) and direct stochastic spin-dynamics (LLL). We have checked that the two approaches are consistent and agree analytically in the linear regime.
Outside the linear regime, we have seen that the product $\om\ta$ is as important as $\xi$ for assessing the validity of LRT. In particular, the perturbative formula Eq.~\eqref{eq:Lambda_3} shows that $\Lambda$ changes sign at $\eta_{c}(\sigma)=\sqrt{(3\sigma-5)/(\sigma+1)}$ ($\simeq1.3$ for $\sigma\simeq5.5$, tending to $\sqrt{3}$ at large $\sigma$), which is distinct from the Debye resonance at $\eta=1$; both thresholds lie close to $T_\mathrm{peak}$, but they are not the same condition. This distinction is made explicit in Fig.~\ref{fig:lambda_eta}.

For superparamagnetic nanomagnets (Fe$_{3}$O$_{4}$, $\gamma$-Fe$_{2}$O$_{3}$) used in MH, the relevant regime is the fast one ($\om\ta\ll1$ at physiological temperature), where LRT \emph{overestimates} the SAR (Sec.~\ref{sec:magnetite}). This is consistent with numerical evidence that LRT overestimates the SAR for small, N\'{e}el-dominated particles at high fields~\cite{Usov2021}. Therefore, protocols based on LRT predictions systematically under-deliver heat, a bias that grows with AC amplitude and with decreasing frequency.
Furthermore, we would like to emphasize that the interplay between N\'{e}el and Brownian relaxation, relevant for nanomagnets in a viscous liquid where both mechanisms contribute, introduces additional nonlinear corrections whose sign and magnitude differ from the purely N\'{e}el case. Accordingly, a direct comparison of LRT with the full nonlinear AC susceptibility in that setting has been carried out experimentally by Yamaminami \emph{et al.}~\cite{Yamaminami2021}. Then, extending the present LLL framework to include rotational degrees of freedom would provide a natural theoretical complement to that work.

On the other hand, the blocked regime ($\om\ta\gg1$), which is accessible to high-anisotropy or large nanomagnets, exhibits the opposite situation. More precisely, LRT underestimates the SAR by up to $\sim70\%$ ($\Lambda\simeq0.7$ at $\xi\simeq2$), as demonstrated for the cobalt model system.

In both cases, the stochastic LLL simulation (or TQMC~\cite{ChubykaloEtAl_prb03, nowaketal00prl, Ledue2026}) provides a more reliable quantitative tool beyond the strict linear regime.

The single-particle picture we developed here is a preliminary study and serves as a controlled benchmark for our ongoing investigation of interacting many-spin assemblies, such as open chains~\cite{Ledue2026}, where
the same LRT-versus-exact comparison can be carried out including dipolar and exchange interactions~\cite{Ruta2015, Dejardin2017, Dejardin2020, Usov2021}.
A limitation of the present work is the assumption of uniaxial magnetocrystalline anisotropy.
In Fe$_{3}$O$_{4}$ and $\gamma$-Fe$_{2}$O$_{3}$ nanoparticles the bulk anisotropy is predominantly cubic, with an effective uniaxial component arising from shape anisotropy and surface effects~\cite{Guardia2012,DeLaPresa2012}; the single effective anisotropy constant $K_{\mathrm{eff}}$ used in Sec.~\ref{sec:magnetite} is a standard approximation that captures the dominant energy scale.
The effect of surface-induced cubic anisotropy on the SAR has in fact already been studied in a companion paper~\cite{MichelsKachkachi2026}, where the effective one-spin problem framework, accounting for both uniaxial and quartic (cubic-symmetry) anisotropy contributions parametrised by $\zeta=K_{4}/K_{2}$, and using the relaxation-time formulae of Refs.~\onlinecite{Dejardin2008,vernayetal14prb}, yields a semi-analytical expression for the specific loss power as a function of~$\zeta$.
The present paper focuses specifically on the LLL-versus-LRT comparison for the uniaxial benchmark case; extending this comparison to the mixed anisotropy setting, building on the results of Ref.~\cite{MichelsKachkachi2026}, is a natural direction for future work.

\section*{Data Availability}
The data that supports the findings of this study are available within the article.

%


\begin{thebibliography}{53}%
\makeatletter
\providecommand \@ifxundefined [1]{%
 \@ifx{#1\undefined}
}%
\providecommand \@ifnum [1]{%
 \ifnum #1\expandafter \@firstoftwo
 \else \expandafter \@secondoftwo
 \fi
}%
\providecommand \@ifx [1]{%
 \ifx #1\expandafter \@firstoftwo
 \else \expandafter \@secondoftwo
 \fi
}%
\providecommand \natexlab [1]{#1}%
\providecommand \enquote  [1]{``#1''}%
\providecommand \bibnamefont  [1]{#1}%
\providecommand \bibfnamefont [1]{#1}%
\providecommand \citenamefont [1]{#1}%
\providecommand \href@noop [0]{\@secondoftwo}%
\providecommand \href [0]{\begingroup \@sanitize@url \@href}%
\providecommand \@href[1]{\@@startlink{#1}\@@href}%
\providecommand \@@href[1]{\endgroup#1\@@endlink}%
\providecommand \@sanitize@url [0]{\catcode `\\12\catcode `\$12\catcode
  `\&12\catcode `\#12\catcode `\^12\catcode `\_12\catcode `\%12\relax}%
\providecommand \@@startlink[1]{}%
\providecommand \@@endlink[0]{}%
\providecommand \url  [0]{\begingroup\@sanitize@url \@url }%
\providecommand \@url [1]{\endgroup\@href {#1}{\urlprefix }}%
\providecommand \urlprefix  [0]{URL }%
\providecommand \Eprint [0]{\href }%
\providecommand \doibase [0]{https://doi.org/}%
\providecommand \selectlanguage [0]{\@gobble}%
\providecommand \bibinfo  [0]{\@secondoftwo}%
\providecommand \bibfield  [0]{\@secondoftwo}%
\providecommand \translation [1]{[#1]}%
\providecommand \BibitemOpen [0]{}%
\providecommand \bibitemStop [0]{}%
\providecommand \bibitemNoStop [0]{.\EOS\space}%
\providecommand \EOS [0]{\spacefactor3000\relax}%
\providecommand \BibitemShut  [1]{\csname bibitem#1\endcsname}%
\let\auto@bib@innerbib\@empty
\bibitem [{\citenamefont {Pankhurst}\ \emph {et~al.}(2003)\citenamefont
  {Pankhurst}, \citenamefont {Connolly}, \citenamefont {Jones},\ and\
  \citenamefont {Dobson}}]{Pankhurst2003}%
  \BibitemOpen
  \bibfield  {author} {\bibinfo {author} {\bibfnamefont {Q.~A.}\ \bibnamefont
  {Pankhurst}}, \bibinfo {author} {\bibfnamefont {J.}~\bibnamefont {Connolly}},
  \bibinfo {author} {\bibfnamefont {S.~K.}\ \bibnamefont {Jones}},\ and\
  \bibinfo {author} {\bibfnamefont {J.}~\bibnamefont {Dobson}},\ }\href
  {https://doi.org/10.1088/0022-3727/36/13/201} {\bibfield  {journal} {\bibinfo
   {journal} {J. Phys. D: Appl. Phys.}\ }\textbf {\bibinfo
  {volume} {36}},\ \bibinfo {pages} {R167} (\bibinfo {year}
  {2003})}\BibitemShut {NoStop}%
\bibitem [{\citenamefont {Dutz}\ and\ \citenamefont {Hergt}(2014)}]{Dutz2014}%
  \BibitemOpen
  \bibfield  {author} {\bibinfo {author} {\bibfnamefont {S.}~\bibnamefont
  {Dutz}}\ and\ \bibinfo {author} {\bibfnamefont {R.}~\bibnamefont {Hergt}},\
  }\href {https://doi.org/10.1088/0957-4484/25/45/452001} {\bibfield  {journal}
  {\bibinfo  {journal} {Nanotechnology}\ }\textbf {\bibinfo {volume} {25}},\
  \bibinfo {pages} {452001} (\bibinfo {year} {2014})}\BibitemShut {NoStop}%
\bibitem [{\citenamefont {{Hergt}}\ \emph {et~al.}(2006)\citenamefont
  {{Hergt}}, \citenamefont {{Dutz}}, \citenamefont {{M{\"u}ller}},\ and\
  \citenamefont {{Zeisberger}}}]{Hergt2006}%
  \BibitemOpen
  \bibfield  {author} {\bibinfo {author} {\bibfnamefont {R.}~\bibnamefont
  {{Hergt}}}, \bibinfo {author} {\bibfnamefont {S.}~\bibnamefont {{Dutz}}},
  \bibinfo {author} {\bibfnamefont {R.}~\bibnamefont {{M{\"u}ller}}},\ and\
  \bibinfo {author} {\bibfnamefont {M.}~\bibnamefont {{Zeisberger}}},\ }\href
  {https://doi.org/10.1088/0953-8984/18/38/S26} {\bibfield  {journal} {\bibinfo
   {journal} {J. Phys.: Condens. Matter}\ }\textbf {\bibinfo {volume} {18}},\
  \bibinfo {pages} {2919} (\bibinfo {year} {2006})}\BibitemShut {NoStop}%
\bibitem [{\citenamefont {Zhang}\ and\ \citenamefont
  {Lu}(2024)}]{Zhang2024MIHReview}%
  \BibitemOpen
  \bibfield  {author} {\bibinfo {author} {\bibfnamefont {Y.-F.}\ \bibnamefont
  {Zhang}}\ and\ \bibinfo {author} {\bibfnamefont {M.}~\bibnamefont {Lu}},\
  }\href {https://doi.org/10.3389/fbioe.2024.1432189} {\bibfield  {journal}
  {\bibinfo  {journal} {Frontiers in Bioengineering and Biotechnology}\
  }\textbf {\bibinfo {volume} {12}},\ \bibinfo {pages} {1432189} (\bibinfo
  {year} {2024})}\BibitemShut {NoStop}%
\bibitem [{\citenamefont {Baldea}\ \emph {et~al.}(2025)\citenamefont {Baldea},
  \citenamefont {Iacovi{\c{t}}{\u{a}}}, \citenamefont {Gurgu}, \citenamefont
  {Vizitiu}, \citenamefont {R{\^a}zniceanu},\ and\ \citenamefont
  {Mitrea}}]{Baldea2025NanoMH}%
  \BibitemOpen
  \bibfield  {author} {\bibinfo {author} {\bibfnamefont {I.}~\bibnamefont
  {Baldea}}, \bibinfo {author} {\bibfnamefont {C.}~\bibnamefont
  {Iacovi{\c{t}}{\u{a}}}}, \bibinfo {author} {\bibfnamefont {R.~A.}\
  \bibnamefont {Gurgu}}, \bibinfo {author} {\bibfnamefont {A.~S.}\ \bibnamefont
  {Vizitiu}}, \bibinfo {author} {\bibfnamefont {V.}~\bibnamefont
  {R{\^a}zniceanu}},\ and\ \bibinfo {author} {\bibfnamefont {D.~R.}\
  \bibnamefont {Mitrea}},\ }\href {https://doi.org/10.3390/nano15191519}
  {\bibfield  {journal} {\bibinfo  {journal} {Nanomaterials}\ }\textbf
  {\bibinfo {volume} {15}},\ \bibinfo {pages} {1519} (\bibinfo {year}
  {2025})}\BibitemShut {NoStop}%
\bibitem [{\citenamefont {Bai}\ \emph {et~al.}(2024)\citenamefont {Bai},
  \citenamefont {Hou}, \citenamefont {Chen}, \citenamefont {Ma}, \citenamefont
  {Gao},\ and\ \citenamefont {Wu}}]{Bai2024MHTTheranostics}%
  \BibitemOpen
  \bibfield  {author} {\bibinfo {author} {\bibfnamefont {S.}~\bibnamefont
  {Bai}}, \bibinfo {author} {\bibfnamefont {S.}~\bibnamefont {Hou}}, \bibinfo
  {author} {\bibfnamefont {T.}~\bibnamefont {Chen}}, \bibinfo {author}
  {\bibfnamefont {X.}~\bibnamefont {Ma}}, \bibinfo {author} {\bibfnamefont
  {C.}~\bibnamefont {Gao}},\ and\ \bibinfo {author} {\bibfnamefont
  {A.}~\bibnamefont {Wu}},\ }\href
  {https://doi.org/10.59717/j.xinn-mater.2024.100051} {\bibfield  {journal}
  {\bibinfo  {journal} {The Innovation Materials}\ }\textbf {\bibinfo {volume}
  {2}},\ \bibinfo {pages} {100051} (\bibinfo {year} {2024})}\BibitemShut
  {NoStop}%
\bibitem [{\citenamefont {D{\'e}jardin}\ and\ \citenamefont
  {Kachkachi}(2022)}]{DejKac2022}%
  \BibitemOpen
  \bibfield  {author} {\bibinfo {author} {\bibfnamefont {J.-L.}\ \bibnamefont
  {D{\'e}jardin}}\ and\ \bibinfo {author} {\bibfnamefont {H.}~\bibnamefont
  {Kachkachi}},\ }\href {https://doi.org/10.1016/j.jmmm.2022.169354} {\bibfield  {journal} {\bibinfo  {journal}
  {J. Magn. Magn. Mater.}\ }\textbf {\bibinfo {volume}
  {556}},\ \bibinfo {pages} {169354} (\bibinfo {year} {2022})}\BibitemShut
  {NoStop}%
\bibitem [{\citenamefont {D{\'e}jardin}\ and\ \citenamefont
  {Kachkachi}(2024)}]{Dejardin2024}%
  \BibitemOpen
  \bibfield  {author} {\bibinfo {author} {\bibfnamefont {J.-L.}\ \bibnamefont
  {D{\'e}jardin}}\ and\ \bibinfo {author} {\bibfnamefont {H.}~\bibnamefont
  {Kachkachi}},\ }\href {https://doi.org/10.3390/app14135757} {\bibfield  {journal} {\bibinfo  {journal}
  {Appl. Sci.}\ }\textbf {\bibinfo {volume} {14}},\ \bibinfo {pages}
  {5757} (\bibinfo {year} {2024})}\BibitemShut {NoStop}%
\bibitem [{\citenamefont {Iglesias}\ \emph {et~al.}(2021)\citenamefont
  {Iglesias} \emph {et~al.}}]{Iglesias2021}%
  \BibitemOpen
  \bibfield  {author} {\bibinfo {author} {\bibfnamefont {C.}~\bibnamefont
  {Iglesias}} \emph {et~al.},\ }\href
  {https://doi.org/10.1038/s41598-021-91282-0} {\bibfield  {journal} {\bibinfo
  {journal} {Sci. Rep.}\ }\textbf {\bibinfo {volume} {11}},\ \bibinfo
  {pages} {11867} (\bibinfo {year} {2021})}\BibitemShut {NoStop}%
\bibitem [{\citenamefont {Ruta}\ \emph {et~al.}(2024)\citenamefont {Ruta},
  \citenamefont {Fern{\'a}ndez-Afonso}, \citenamefont {Rannala}, \citenamefont
  {Morales}, \citenamefont {Veintemillas-Verdaguer}, \citenamefont {Jones},
  \citenamefont {Guti{\'e}rrez}, \citenamefont {Chantrell},\ and\ \citenamefont
  {Serantes}}]{Ruta2024}%
  \BibitemOpen
  \bibfield  {author} {\bibinfo {author} {\bibfnamefont {S.}~\bibnamefont
  {Ruta}}, \bibinfo {author} {\bibfnamefont {Y.}~\bibnamefont
  {Fern{\'a}ndez-Afonso}}, \bibinfo {author} {\bibfnamefont {S.~E.}\
  \bibnamefont {Rannala}}, \bibinfo {author} {\bibfnamefont {M.~P.}\
  \bibnamefont {Morales}}, \bibinfo {author} {\bibfnamefont {S.}~\bibnamefont
  {Veintemillas-Verdaguer}}, \bibinfo {author} {\bibfnamefont {C.}~\bibnamefont
  {Jones}}, \bibinfo {author} {\bibfnamefont {L.}~\bibnamefont
  {Guti{\'e}rrez}}, \bibinfo {author} {\bibfnamefont {R.~W.}\ \bibnamefont
  {Chantrell}},\ and\ \bibinfo {author} {\bibfnamefont {D.}~\bibnamefont
  {Serantes}},\ }\href@noop {} {\bibfield  {journal} {\bibinfo  {journal}
  {Nanoscale Advances}\ }\textbf {\bibinfo {volume} {6}},\ \bibinfo {pages}
  {4207} (\bibinfo {year} {2024})}\BibitemShut {NoStop}%
\bibitem [{\citenamefont {Kachkachi}(2026)}]{Kachkachi2026LocalGlobalHeating}%
  \BibitemOpen
  \bibfield  {author} {\bibinfo {author} {\bibfnamefont {H.}~\bibnamefont
  {Kachkachi}},\ }\href {https://arxiv.org/abs/2606.04724} {\enquote {\bibinfo
  {title} {Local-to-global heating crossover in chains of nanomagnets: A
  two-scale analytical framework},}\ } (\bibinfo {year} {2026}),\ \bibinfo
  {note} {submitted for publication; arXiv:2606.04724v1},\ \Eprint
  {https://arxiv.org/abs/2606.04724} {arXiv:2606.04724 [cond-mat.mes-hall]}
  \BibitemShut {NoStop}%
\bibitem [{\citenamefont {Rosensweig}(2002)}]{Rosensweig2002}%
  \BibitemOpen
  \bibfield  {author} {\bibinfo {author} {\bibfnamefont {R.~E.}\ \bibnamefont
  {Rosensweig}},\ }\href
  {http://www.sciencedirect.com/science/article/pii/S0304885302007060}
  {\bibfield  {journal} {\bibinfo  {journal} {J. Magn. Magn. Mater.}\ }\textbf
  {\bibinfo {volume} {252}},\ \bibinfo {pages} {370} (\bibinfo {year}
  {2002})},\ \bibinfo {note} {proceedings of the 9th International Conference
  on Magnetic Fluids, 23-27 Jul. 2001}\BibitemShut {NoStop}%
\bibitem [{\citenamefont {D{\'e}jardin}\ \emph {et~al.}(2017)\citenamefont
  {D{\'e}jardin}, \citenamefont {Vernay}, \citenamefont {Respaud},\ and\
  \citenamefont {Kachkachi}}]{Dejardin2017}%
  \BibitemOpen
  \bibfield  {author} {\bibinfo {author} {\bibfnamefont {J.-L.}\ \bibnamefont
  {D{\'e}jardin}}, \bibinfo {author} {\bibfnamefont {F.}~\bibnamefont
  {Vernay}}, \bibinfo {author} {\bibfnamefont {M.}~\bibnamefont {Respaud}},\
  and\ \bibinfo {author} {\bibfnamefont {H.}~\bibnamefont {Kachkachi}},\
  }\href {https://doi.org/10.1063/1.4983297} {\bibfield  {journal} {\bibinfo  {journal} {J. Appl. Phys.}\
  }\textbf {\bibinfo {volume} {121}},\ \bibinfo {pages} {203903} (\bibinfo
  {year} {2017})}\BibitemShut {NoStop}%
\bibitem [{\citenamefont {N{\'e}el}(1949)}]{Neel1949}%
  \BibitemOpen
  \bibfield  {author} {\bibinfo {author} {\bibfnamefont {L.}~\bibnamefont
  {N{\'e}el}},\ }\href@noop {} {\bibfield  {journal} {\bibinfo  {journal}
  {Annales de G{\'e}ophysique}\ }\textbf {\bibinfo {volume} {5}},\ \bibinfo
  {pages} {99} (\bibinfo {year} {1949})}\BibitemShut {NoStop}%
\bibitem [{\citenamefont {Brown}(1963)}]{Brown1963}%
  \BibitemOpen
  \bibfield  {author} {\bibinfo {author} {\bibfnamefont {W.~F.}\ \bibnamefont
  {Brown}},\ }\href {https://doi.org/10.1103/PhysRev.130.1677} {\bibfield
  {journal} {\bibinfo  {journal} {Physical Review}\ }\textbf {\bibinfo {volume}
  {130}},\ \bibinfo {pages} {1677} (\bibinfo {year} {1963})}\BibitemShut
  {NoStop}%
\bibitem [{\citenamefont {Carrey}, \citenamefont {Mehdaoui},\ and\
  \citenamefont {Respaud}(2011)}]{Carrey2011}%
  \BibitemOpen
  \bibfield  {author} {\bibinfo {author} {\bibfnamefont {J.}~\bibnamefont
  {Carrey}}, \bibinfo {author} {\bibfnamefont {B.}~\bibnamefont {Mehdaoui}},\
  and\ \bibinfo {author} {\bibfnamefont {M.}~\bibnamefont {Respaud}},\
  }\href {https://doi.org/10.1063/1.3551496} {\bibfield  {journal} {\bibinfo  {journal} {J. Appl. Phys.}\
  }\textbf {\bibinfo {volume} {109}},\ \bibinfo {pages} {083912} (\bibinfo
  {year} {2011})}\BibitemShut {NoStop}%
\bibitem [{\citenamefont {Andreu}\ and\ \citenamefont
  {Natividad}(2013)}]{Andreu2013}%
  \BibitemOpen
  \bibfield  {author} {\bibinfo {author} {\bibfnamefont {I.}~\bibnamefont
  {Andreu}}\ and\ \bibinfo {author} {\bibfnamefont {E.}~\bibnamefont
  {Natividad}},\ }\href {https://doi.org/10.3109/02656736.2013.826825}
  {\bibfield  {journal} {\bibinfo  {journal} {International Journal of
  Hyperthermia}\ }\textbf {\bibinfo {volume} {29}},\ \bibinfo {pages} {739}
  (\bibinfo {year} {2013})}\BibitemShut {NoStop}%
\bibitem [{\citenamefont {Usov}, \citenamefont {Rytov},\ and\ \citenamefont
  {Bautin}(2021)}]{Usov2021}%
  \BibitemOpen
  \bibfield  {author} {\bibinfo {author} {\bibfnamefont {N.~A.}\ \bibnamefont
  {Usov}}, \bibinfo {author} {\bibfnamefont {R.~A.}\ \bibnamefont {Rytov}},\
  and\ \bibinfo {author} {\bibfnamefont {V.~A.}\ \bibnamefont {Bautin}},\
  }\href {https://doi.org/10.1038/s41598-021-86323-x} {\bibfield  {journal}
  {\bibinfo  {journal} {Sci. Rep.}\ }\textbf {\bibinfo {volume}
  {11}},\ \bibinfo {pages} {6999} (\bibinfo {year} {2021})}\BibitemShut
  {NoStop}%
\bibitem [{\citenamefont {Yamaminami}\ \emph {et~al.}(2021)\citenamefont
  {Yamaminami}, \citenamefont {Ota}, \citenamefont {Trisnanto}, \citenamefont
  {Ishikawa}, \citenamefont {Yamada}, \citenamefont {Yoshida}, \citenamefont
  {Enpuku},\ and\ \citenamefont {Takemura}}]{Yamaminami2021}%
  \BibitemOpen
  \bibfield  {author} {\bibinfo {author} {\bibfnamefont {T.}~\bibnamefont
  {Yamaminami}}, \bibinfo {author} {\bibfnamefont {S.}~\bibnamefont {Ota}},
  \bibinfo {author} {\bibfnamefont {S.~B.}\ \bibnamefont {Trisnanto}}, \bibinfo
  {author} {\bibfnamefont {M.}~\bibnamefont {Ishikawa}}, \bibinfo {author}
  {\bibfnamefont {T.}~\bibnamefont {Yamada}}, \bibinfo {author} {\bibfnamefont
  {T.}~\bibnamefont {Yoshida}}, \bibinfo {author} {\bibfnamefont
  {K.}~\bibnamefont {Enpuku}},\ and\ \bibinfo {author} {\bibfnamefont
  {Y.}~\bibnamefont {Takemura}},\ }\href
  {https://doi.org/10.1016/j.jmmm.2020.167401} {\bibfield  {journal} {\bibinfo
  {journal} {J. Magn. Magn. Mater.}\ }\textbf {\bibinfo
  {volume} {517}},\ \bibinfo {pages} {167401} (\bibinfo {year}
  {2021})}\BibitemShut {NoStop}%
\bibitem [{\citenamefont {Lacroix}\ \emph {et~al.}(2009)\citenamefont
  {Lacroix}, \citenamefont {Malaki}, \citenamefont {Carrey}, \citenamefont
  {Lachaize}, \citenamefont {Respaud}, \citenamefont {Goya},\ and\
  \citenamefont {Chaudret}}]{Lacroix_etal_JAP2009}%
  \BibitemOpen
  \bibfield  {author} {\bibinfo {author} {\bibfnamefont {L.-M.}\ \bibnamefont
  {Lacroix}}, \bibinfo {author} {\bibfnamefont {R.~B.}\ \bibnamefont {Malaki}},
  \bibinfo {author} {\bibfnamefont {J.}~\bibnamefont {Carrey}}, \bibinfo
  {author} {\bibfnamefont {S.}~\bibnamefont {Lachaize}}, \bibinfo {author}
  {\bibfnamefont {M.}~\bibnamefont {Respaud}}, \bibinfo {author} {\bibfnamefont
  {G.~F.}\ \bibnamefont {Goya}},\ and\ \bibinfo {author} {\bibfnamefont
  {B.}~\bibnamefont {Chaudret}},\ }\href {https://doi.org/10.1063/1.3068195}
  {\bibfield  {journal} {\bibinfo  {journal} {J. Appl. Phys.}\ }\textbf
  {\bibinfo {volume} {105}},\ \bibinfo {pages} {023911} (\bibinfo {year}
  {2009})}\BibitemShut {NoStop}%
\bibitem [{\citenamefont {Martinez-Boubeta}\ \emph {et~al.}(2013)\citenamefont
  {Martinez-Boubeta}, \citenamefont {Simeonidis}, \citenamefont {Makridis},
  \citenamefont {Angelakeris}, \citenamefont {Iglesias}, \citenamefont
  {Guardia}, \citenamefont {Cabot}, \citenamefont {Yedra}, \citenamefont
  {Estrad{\'e}}, \citenamefont {Peir{\'o}} \emph
  {et~al.}}]{martinez2013learning}%
  \BibitemOpen
  \bibfield  {author} {\bibinfo {author} {\bibfnamefont {C.}~\bibnamefont
  {Martinez-Boubeta}}, \bibinfo {author} {\bibfnamefont {K.}~\bibnamefont
  {Simeonidis}}, \bibinfo {author} {\bibfnamefont {A.}~\bibnamefont
  {Makridis}}, \bibinfo {author} {\bibfnamefont {M.}~\bibnamefont
  {Angelakeris}}, \bibinfo {author} {\bibfnamefont {O.}~\bibnamefont
  {Iglesias}}, \bibinfo {author} {\bibfnamefont {P.}~\bibnamefont {Guardia}},
  \bibinfo {author} {\bibfnamefont {A.}~\bibnamefont {Cabot}}, \bibinfo
  {author} {\bibfnamefont {L.}~\bibnamefont {Yedra}}, \bibinfo {author}
  {\bibfnamefont {S.}~\bibnamefont {Estrad{\'e}}}, \bibinfo {author}
  {\bibfnamefont {F.}~\bibnamefont {Peir{\'o}}}, \emph {et~al.},\ }\href
  {http://dx.doi.org/10.1038/srep01652} {\bibfield  {journal} {\bibinfo
  {journal} {Sci. Rep.}\ }\textbf {\bibinfo {volume} {3}},\ \bibinfo {pages}
  {1652} (\bibinfo {year} {2013})}\BibitemShut {NoStop}%
\bibitem [{\citenamefont {Mehdaoui}\ \emph {et~al.}(2013)\citenamefont
  {Mehdaoui}, \citenamefont {Tan}, \citenamefont {Meffre}, \citenamefont
  {Carrey}, \citenamefont {Lachaize}, \citenamefont {Chaudret},\ and\
  \citenamefont {Respaud}}]{Mehdaoui_prb2013}%
  \BibitemOpen
  \bibfield  {author} {\bibinfo {author} {\bibfnamefont {B.}~\bibnamefont
  {Mehdaoui}}, \bibinfo {author} {\bibfnamefont {R.~P.}\ \bibnamefont {Tan}},
  \bibinfo {author} {\bibfnamefont {A.}~\bibnamefont {Meffre}}, \bibinfo
  {author} {\bibfnamefont {J.}~\bibnamefont {Carrey}}, \bibinfo {author}
  {\bibfnamefont {S.}~\bibnamefont {Lachaize}}, \bibinfo {author}
  {\bibfnamefont {B.}~\bibnamefont {Chaudret}},\ and\ \bibinfo {author}
  {\bibfnamefont {M.}~\bibnamefont {Respaud}},\ }\href
  {https://doi.org/10.1103/PhysRevB.87.174419} {\bibfield  {journal} {\bibinfo
  {journal} {Phys. Rev. B}\ }\textbf {\bibinfo {volume} {87}},\ \bibinfo
  {pages} {174419} (\bibinfo {year} {2013})}\BibitemShut {NoStop}%
\bibitem [{\citenamefont {Chubykalo}\ \emph {et~al.}(2003)\citenamefont
  {Chubykalo}, \citenamefont {Nowak}, \citenamefont {Smirnov-Rueda},
  \citenamefont {Wongsam}, \citenamefont {Chantrell},\ and\ \citenamefont
  {Gonzalez}}]{ChubykaloEtAl_prb03}%
  \BibitemOpen
  \bibfield  {author} {\bibinfo {author} {\bibfnamefont {O.}~\bibnamefont
  {Chubykalo}}, \bibinfo {author} {\bibfnamefont {U.}~\bibnamefont {Nowak}},
  \bibinfo {author} {\bibfnamefont {R.}~\bibnamefont {Smirnov-Rueda}},
  \bibinfo {author} {\bibfnamefont {M.~A.}\ \bibnamefont {Wongsam}},
  \bibinfo {author} {\bibfnamefont {R.~W.}\ \bibnamefont {Chantrell}},\ and\
  \bibinfo {author} {\bibfnamefont {J.~M.}\ \bibnamefont {Gonzalez}},\
  }\href {https://doi.org/10.1103/PhysRevB.67.064422} {\bibfield  {journal} {\bibinfo  {journal} {Phys. Rev. B}\
  }\textbf {\bibinfo {volume} {67}},\ \bibinfo {pages} {064422} (\bibinfo
  {year} {2003})}\BibitemShut {NoStop}%
\bibitem [{\citenamefont {Nowak}, \citenamefont {Chantrell},\ and\ \citenamefont
  {Kennedy}(2000)}]{nowaketal00prl}%
  \BibitemOpen
  \bibfield  {author} {\bibinfo {author} {\bibfnamefont {U.}~\bibnamefont
  {Nowak}}, \bibinfo {author} {\bibfnamefont {R.~W.}\ \bibnamefont {Chantrell}},\
  and\ \bibinfo {author} {\bibfnamefont {E.~C.}\ \bibnamefont {Kennedy}},\
  }\href {https://doi.org/10.1103/PhysRevLett.83.163} {\bibfield  {journal}
  {\bibinfo  {journal} {Phys. Rev. Lett.}\ }\textbf {\bibinfo {volume}
  {83}},\ \bibinfo {pages} {163} (\bibinfo {year} {2000})}\BibitemShut
  {NoStop}%
\bibitem [{\citenamefont {Ledue}, \citenamefont {Vernay},\ and\ \citenamefont
  {Kachkachi}(2026)}]{Ledue2026}%
  \BibitemOpen
  \bibfield  {author} {\bibinfo {author} {\bibfnamefont {D.}~\bibnamefont
  {Ledue}}, \bibinfo {author} {\bibfnamefont {F.}~\bibnamefont {Vernay}},\ and\
  \bibinfo {author} {\bibfnamefont {H.}~\bibnamefont {Kachkachi}},\ }\href@noop
  {} {\bibfield  {journal} {\bibinfo  {journal} {J. Magn. Magn. Mater.}\ }
  (\bibinfo {year} {2026})},\ \bibinfo {note} {in press; \href {https://doi.org/10.48550/arXiv.2609.39811} {arXiv:2609.39811}}\BibitemShut {NoStop}%
\bibitem [{\citenamefont {Brown}(1979)}]{Brown1963stoch}%
  \BibitemOpen
  \bibfield  {author} {\bibinfo {author} {\bibfnamefont {W.}~\bibnamefont
  {Brown}},\ }\href {https://doi.org/10.1109/TMAG.1979.1060329} {\bibfield
  {journal} {\bibinfo  {journal} {IEEE Transactions on Magnetics}\ }\textbf
  {\bibinfo {volume} {15}},\ \bibinfo {pages} {1196} (\bibinfo {year}
  {1979})}\BibitemShut {NoStop}%
\bibitem [{\citenamefont {Garc{\'i}a-Palacios}\ and\ \citenamefont
  {L{\'a}zaro}(1998)}]{GarciaPalacios1998}%
  \BibitemOpen
  \bibfield  {author} {\bibinfo {author} {\bibfnamefont {J.~L.}\ \bibnamefont
  {Garc{\'i}a-Palacios}}\ and\ \bibinfo {author} {\bibfnamefont {F.~J.}\
  \bibnamefont {L{\'a}zaro}},\ }\href
  {https://doi.org/10.1103/PhysRevB.58.14937} {\bibfield  {journal} {\bibinfo
  {journal} {Phys. Rev. B}\ }\textbf {\bibinfo {volume} {58}},\ \bibinfo
  {pages} {14937} (\bibinfo {year} {1998})}\BibitemShut {NoStop}%
\bibitem [{\citenamefont {Garc{\'i}a-Palacios}(2007)}]{garpal00acp}%
  \BibitemOpen
  \bibfield  {author} {\bibinfo {author} {\bibfnamefont {J.~L.}\ \bibnamefont
  {Garc{\'i}a-Palacios}},\ }\enquote {\bibinfo {title} {{On the Statics and
  Dynamics of Magnetoanisotropic Nanoparticles}},}\ in\ \href
  {https://doi.org/10.1002/9780470141717.ch1} {\emph {\bibinfo {booktitle}
  {{Advances in Chemical Physics}}}},\ Vol.\ \bibinfo {volume} {112}\ (\bibinfo
   {publisher} {John Wiley \& Sons, Inc.},\ \bibinfo {year} {2007})\ pp.\
  \bibinfo {pages} {1--210}\BibitemShut {NoStop}%
\bibitem [{\citenamefont {D{\'e}jardin}, \citenamefont {Kachkachi},\ and\
  \citenamefont {Martinez}(2012)}]{DejardinKachkachiMartinez2012Comment}%
  \BibitemOpen
  \bibfield  {author} {\bibinfo {author} {\bibfnamefont {J.-L.}\ \bibnamefont
  {D{\'e}jardin}}, \bibinfo {author} {\bibfnamefont {H.}~\bibnamefont
  {Kachkachi}},\ and\ \bibinfo {author} {\bibfnamefont {J.-M.}\ \bibnamefont
  {Martinez}},\ }\href {https://arxiv.org/abs/1210.2436} {\enquote {\bibinfo
  {title} {Comment on ``thermal fluctuations of magnetic
  nanoparticles''[arxiv:1209.0298]},}\ } (\bibinfo {year} {2012}),\ \Eprint
  {https://arxiv.org/abs/1210.2436} {arXiv:1210.2436 [cond-mat.mes-hall]}
  \BibitemShut {NoStop}%
\bibitem [{\citenamefont {Hammersley}\ and\ \citenamefont
  {Morton}(1956)}]{Hammersley1956}%
  \BibitemOpen
  \bibfield  {author} {\bibinfo {author} {\bibfnamefont {J.~M.}\ \bibnamefont
  {Hammersley}}\ and\ \bibinfo {author} {\bibfnamefont {K.~W.}\ \bibnamefont
  {Morton}},\ }\href {https://doi.org/10.1017/s0305004100031455} {\bibfield
  {journal} {\bibinfo  {journal} {Mathematical Proceedings of the Cambridge
  Philosophical Society}\ }\textbf {\bibinfo {volume} {52}},\ \bibinfo {pages}
  {449} (\bibinfo {year} {1956})}\BibitemShut {NoStop}%
\bibitem [{\citenamefont {Owen}(2008)}]{Owen2008}%
  \BibitemOpen
  \bibfield  {author} {\bibinfo {author} {\bibfnamefont {A.~B.}\ \bibnamefont
  {Owen}},\ }\href {https://doi.org/10.1214/07-aos548} {\bibfield  {journal}
  {\bibinfo  {journal} {The Annals of Statistics}\ }\textbf {\bibinfo {volume}
  {36}} (\bibinfo {year} {2008}),\ 10.1214/07-aos548}\BibitemShut {NoStop}%
\bibitem [{\citenamefont {Kubo}, \citenamefont {Toda},\ and\ \citenamefont
  {Hashitsume}(1991)}]{KuboTodaHashitsume1991}%
  \BibitemOpen
  \bibfield  {author} {\bibinfo {author} {\bibfnamefont {R.}~\bibnamefont
  {Kubo}}, \bibinfo {author} {\bibfnamefont {M.}~\bibnamefont {Toda}},\ and\
  \bibinfo {author} {\bibfnamefont {N.}~\bibnamefont {Hashitsume}},\ }\href
  {https://doi.org/10.1007/978-3-642-58244-8} {\emph {\bibinfo {title}
  {Statistical Physics II: Nonequilibrium Statistical Mechanics}}},\ \bibinfo
  {series} {Springer Series in Solid-State Sciences}, Vol.~\bibinfo {volume}
  {31}\ (\bibinfo  {publisher} {Springer},\ \bibinfo {address} {Berlin,
  Heidelberg},\ \bibinfo {year} {1991})\BibitemShut {NoStop}%
\bibitem [{\citenamefont {Coffey}\ \emph {et~al.}(1995)\citenamefont {Coffey},
  \citenamefont {Crothers}, \citenamefont {Kalmykov},\ and\ \citenamefont
  {Waldron}}]{CoffeyEtal_PhysRevB.51.15947}%
  \BibitemOpen
  \bibfield  {author} {\bibinfo {author} {\bibfnamefont {W.~T.}\ \bibnamefont
  {Coffey}}, \bibinfo {author} {\bibfnamefont {D.~S.~F.}\ \bibnamefont
  {Crothers}}, \bibinfo {author} {\bibfnamefont {Y.~P.}\ \bibnamefont
  {Kalmykov}},\ and\ \bibinfo {author} {\bibfnamefont {J.~T.}\ \bibnamefont
  {Waldron}},\ }\href {https://doi.org/10.1103/PhysRevB.51.15947} {\bibfield
  {journal} {\bibinfo  {journal} {Phys. Rev. B}\ }\textbf {\bibinfo {volume}
  {51}},\ \bibinfo {pages} {15947} (\bibinfo {year} {1995})}\BibitemShut
  {NoStop}%
\bibitem [{\citenamefont {Cregg}, \citenamefont {Crothers},\ and\ \citenamefont
  {Wickstead}(1994)}]{creetal94jap}%
  \BibitemOpen
  \bibfield  {author} {\bibinfo {author} {\bibfnamefont {P.~J.}\ \bibnamefont
  {Cregg}}, \bibinfo {author} {\bibfnamefont {D.~S.~F.}\ \bibnamefont
  {Crothers}},\ and\ \bibinfo {author} {\bibfnamefont {A.~W.}\ \bibnamefont
  {Wickstead}},\ }\href {https://doi.org/10.1063/1.357340} {\bibfield  {journal}
  {\bibinfo  {journal} {J. Appl. Phys.}\ }\textbf {\bibinfo {volume} {76}},\
  \bibinfo {pages} {4900} (\bibinfo {year} {1994})}\BibitemShut {NoStop}%
\bibitem [{\citenamefont {Bitoh}\ \emph {et~al.}(1995)\citenamefont {Bitoh},
  \citenamefont {Ohba}, \citenamefont {Takamatsu}, \citenamefont {Shirane},\
  and\ \citenamefont {Chikazawa}}]{BitohEtal_JPSJ.64.1311}%
  \BibitemOpen
  \bibfield  {author} {\bibinfo {author} {\bibfnamefont {T.}~\bibnamefont
  {Bitoh}}, \bibinfo {author} {\bibfnamefont {K.}~\bibnamefont {Ohba}},
  \bibinfo {author} {\bibfnamefont {M.}~\bibnamefont {Takamatsu}}, \bibinfo
  {author} {\bibfnamefont {T.}~\bibnamefont {Shirane}},\ and\ \bibinfo {author}
  {\bibfnamefont {S.}~\bibnamefont {Chikazawa}},\ }\href
  {https://doi.org/10.1143/JPSJ.64.1311} {\bibfield  {journal} {\bibinfo
  {journal} {Journal of the Physical Society of Japan}\ }\textbf {\bibinfo
  {volume} {64}},\ \bibinfo {pages} {1311} (\bibinfo {year} {1995})}\BibitemShut {NoStop}%
\bibitem [{\citenamefont {Bitoh}\ \emph {et~al.}(1996)\citenamefont {Bitoh},
  \citenamefont {Ohba}, \citenamefont {Takamatsu}, \citenamefont {Shirane},\
  and\ \citenamefont {Chikazawa}}]{BitohEtal_J3M.154.59}%
  \BibitemOpen
  \bibfield  {author} {\bibinfo {author} {\bibfnamefont {T.}~\bibnamefont
  {Bitoh}}, \bibinfo {author} {\bibfnamefont {K.}~\bibnamefont {Ohba}},
  \bibinfo {author} {\bibfnamefont {M.}~\bibnamefont {Takamatsu}}, \bibinfo
  {author} {\bibfnamefont {T.}~\bibnamefont {Shirane}},\ and\ \bibinfo {author}
  {\bibfnamefont {S.}~\bibnamefont {Chikazawa}},\ }\href
  {https://doi.org/10.1016/0304-8853(95)00572-2} {\bibfield  {journal}
  {\bibinfo  {journal} {J. Magn. Magn. Mater.}\ }\textbf {\bibinfo {volume}
  {154}},\ \bibinfo {pages} {59} (\bibinfo {year} {1996})}\BibitemShut
  {NoStop}%
\bibitem [{\citenamefont {Coffey}\ \emph {et~al.}(1998)\citenamefont {Coffey},
  \citenamefont {Crothers}, \citenamefont {Dormann}, \citenamefont {Kalmykov},
  \citenamefont {Kennedy},\ and\ \citenamefont {Wernsdorfer}}]{Coffey1998}%
  \BibitemOpen
  \bibfield  {author} {\bibinfo {author} {\bibfnamefont {W.~T.}\ \bibnamefont
  {Coffey}}, \bibinfo {author} {\bibfnamefont {D.~S.~F.}\ \bibnamefont
  {Crothers}}, \bibinfo {author} {\bibfnamefont {J.~L.}\ \bibnamefont
  {Dormann}}, \bibinfo {author} {\bibfnamefont {Y.~P.}\ \bibnamefont
  {Kalmykov}}, \bibinfo {author} {\bibfnamefont {E.~C.}\ \bibnamefont
  {Kennedy}},\ and\ \bibinfo {author} {\bibfnamefont {W.}~\bibnamefont
  {Wernsdorfer}},\ }\href {https://doi.org/10.1103/PhysRevLett.80.5655}
  {\bibfield  {journal} {\bibinfo  {journal} {Phys. Rev. Lett.}\
  }\textbf {\bibinfo {volume} {80}},\ \bibinfo {pages} {5655} (\bibinfo {year}
  {1998})}\BibitemShut {NoStop}%
\bibitem [{\citenamefont {Vernay}, \citenamefont {Sabsabi},\ and\ \citenamefont
  {Kachkachi}(2014)}]{vernayetal14prb}%
  \BibitemOpen
  \bibfield  {author} {\bibinfo {author} {\bibfnamefont {F.}~\bibnamefont
  {Vernay}}, \bibinfo {author} {\bibfnamefont {Z.}~\bibnamefont {Sabsabi}},\
  and\ \bibinfo {author} {\bibfnamefont {H.}~\bibnamefont {Kachkachi}},\
  }\href {https://doi.org/10.1103/PhysRevB.90.094416} {\bibfield  {journal} {\bibinfo  {journal}
  {Phys. Rev. B}\ }\textbf {\bibinfo {volume} {90}},\ \bibinfo {pages}
  {094416} (\bibinfo {year} {2014})}\BibitemShut {NoStop}%
\bibitem [{\citenamefont {Garc{\'\i}a-Palacios}\ and\ \citenamefont
  {Garanin}(2004)}]{garpalgar04prb}%
  \BibitemOpen
  \bibfield  {author} {\bibinfo {author} {\bibfnamefont {J.~L.}\ \bibnamefont
  {Garc{\'\i}a-Palacios}}\ and\ \bibinfo {author} {\bibfnamefont {D.~A.}\
  \bibnamefont {Garanin}},\ }\href {https://doi.org/10.1103/PhysRevB.70.064415}
  {\bibfield  {journal} {\bibinfo  {journal} {Phys. Rev. B}\ }\textbf {\bibinfo
  {volume} {70}},\ \bibinfo {pages} {064415} (\bibinfo {year}
  {2004})}\BibitemShut {NoStop}%
\bibitem [{\citenamefont {J{\"o}nsson}\ and\ \citenamefont
  {Garc{\'\i}a-Palacios}(2001)}]{JonssonGarciaPalacios2001}%
  \BibitemOpen
  \bibfield  {author} {\bibinfo {author} {\bibfnamefont {P.~E.}\ \bibnamefont
  {J{\"o}nsson}}\ and\ \bibinfo {author} {\bibfnamefont {J.~L.}\ \bibnamefont
  {Garc{\'\i}a-Palacios}},\ }\bibfield  {journal} {\bibinfo  {journal}
  {Europhys. Lett.}\ }\textbf {\bibinfo {volume} {55}},\ \bibinfo {pages} {418}
  (\bibinfo {year} {2001})\BibitemShut {NoStop}%
\bibitem [{\citenamefont {D{\'e}jardin}, \citenamefont {Vernay},\ and\
  \citenamefont {Kachkachi}(2020)}]{Dejardin2020}%
  \BibitemOpen
  \bibfield  {author} {\bibinfo {author} {\bibfnamefont {J.-L.}\ \bibnamefont
  {D{\'e}jardin}}, \bibinfo {author} {\bibfnamefont {F.}~\bibnamefont
  {Vernay}},\ and\ \bibinfo {author} {\bibfnamefont {H.}~\bibnamefont
  {Kachkachi}},\ }\href {https://doi.org/10.1063/5.0011888} {\bibfield  {journal} {\bibinfo  {journal} {J.
  Appl. Phys.}\ }\textbf {\bibinfo {volume} {128}},\ \bibinfo {pages} {143901}
  (\bibinfo {year} {2020})}\BibitemShut {NoStop}%
\bibitem [{\citenamefont {Jamet}\ \emph {et~al.}(2001)\citenamefont {Jamet},
  \citenamefont {Wernsdorfer}, \citenamefont {Thirion}, \citenamefont {Mailly},
  \citenamefont {Dupuis}, \citenamefont {M{\'e}linon},\ and\ \citenamefont
  {P{\'e}rez}}]{jametetal01prl}%
  \BibitemOpen
  \bibfield  {author} {\bibinfo {author} {\bibfnamefont {M.}~\bibnamefont
  {Jamet}}, \bibinfo {author} {\bibfnamefont {W.}~\bibnamefont {Wernsdorfer}},
  \bibinfo {author} {\bibfnamefont {C.}~\bibnamefont {Thirion}}, \bibinfo
  {author} {\bibfnamefont {D.}~\bibnamefont {Mailly}}, \bibinfo {author}
  {\bibfnamefont {V.}~\bibnamefont {Dupuis}}, \bibinfo {author} {\bibfnamefont
  {P.}~\bibnamefont {M{\'e}linon}},\ and\ \bibinfo {author} {\bibfnamefont
  {A.}~\bibnamefont {P{\'e}rez}},\ }\href {https://doi.org/10.1103/PhysRevLett.86.4676} {\bibfield  {journal} {\bibinfo  {journal}
  {Phys. Rev. Lett.}\ }\textbf {\bibinfo {volume} {86}},\ \bibinfo {pages}
  {4676} (\bibinfo {year} {2001})}\BibitemShut {NoStop}%
\bibitem [{\citenamefont {Jamet}\ \emph {et~al.}(2004)\citenamefont {Jamet},
  \citenamefont {Wernsdorfer}, \citenamefont {Thirion}, \citenamefont {Dupuis},
  \citenamefont {M{\'e}linon}, \citenamefont {P{\'e}rez},\ and\ \citenamefont
  {Mailly}}]{jametetal04prb}%
  \BibitemOpen
  \bibfield  {author} {\bibinfo {author} {\bibfnamefont {M.}~\bibnamefont
  {Jamet}}, \bibinfo {author} {\bibfnamefont {W.}~\bibnamefont {Wernsdorfer}},
  \bibinfo {author} {\bibfnamefont {C.}~\bibnamefont {Thirion}}, \bibinfo
  {author} {\bibfnamefont {V.}~\bibnamefont {Dupuis}}, \bibinfo {author}
  {\bibfnamefont {P.}~\bibnamefont {M{\'e}linon}}, \bibinfo {author}
  {\bibfnamefont {A.}~\bibnamefont {P{\'e}rez}},\ and\ \bibinfo {author}
  {\bibfnamefont {D.}~\bibnamefont {Mailly}},\ }\href {https://doi.org/10.1103/PhysRevB.69.024401} {\bibfield  {journal} {\bibinfo  {journal} {Phys.
  Rev. B}\ }\textbf {\bibinfo {volume} {69}},\ \bibinfo {pages} {024401}
  (\bibinfo {year} {2004})}\BibitemShut {NoStop}%
\bibitem [{\citenamefont {Fortin}\ \emph {et~al.}(2007)\citenamefont {Fortin},
  \citenamefont {Wilhelm}, \citenamefont {Servais}, \citenamefont
  {M{\'e}nager}, \citenamefont {Bacri},\ and\ \citenamefont
  {Gazeau}}]{Fortin2007}%
  \BibitemOpen
  \bibfield  {author} {\bibinfo {author} {\bibfnamefont {J.-P.}\ \bibnamefont
  {Fortin}}, \bibinfo {author} {\bibfnamefont {C.}~\bibnamefont {Wilhelm}},
  \bibinfo {author} {\bibfnamefont {J.}~\bibnamefont {Servais}}, \bibinfo
  {author} {\bibfnamefont {C.}~\bibnamefont {M{\'e}nager}}, \bibinfo {author}
  {\bibfnamefont {J.-C.}\ \bibnamefont {Bacri}},\ and\ \bibinfo {author}
  {\bibfnamefont {F.}~\bibnamefont {Gazeau}},\ }\href
  {https://doi.org/10.1021/ja067457e} {\bibfield  {journal} {\bibinfo
  {journal} {Journal of the American Chemical Society}\ }\textbf {\bibinfo
  {volume} {129}},\ \bibinfo {pages} {2628} (\bibinfo {year}
  {2007})}\BibitemShut {NoStop}%
\bibitem [{\citenamefont {Serantes}\ \emph {et~al.}(2018)\citenamefont
  {Serantes}, \citenamefont {Chantrell}, \citenamefont {Gavil{\'a}n},
  \citenamefont {Morales}, \citenamefont {Chubykalo-Fesenko}, \citenamefont
  {Baldomir},\ and\ \citenamefont {Satoh}}]{Serantes2018}%
  \BibitemOpen
  \bibfield  {author} {\bibinfo {author} {\bibfnamefont {D.}~\bibnamefont
  {Serantes}}, \bibinfo {author} {\bibfnamefont {R.}~\bibnamefont {Chantrell}},
  \bibinfo {author} {\bibfnamefont {H.}~\bibnamefont {Gavil{\'a}n}}, \bibinfo
  {author} {\bibfnamefont {M.~d.~P.}\ \bibnamefont {Morales}}, \bibinfo
  {author} {\bibfnamefont {O.}~\bibnamefont {Chubykalo-Fesenko}}, \bibinfo
  {author} {\bibfnamefont {D.}~\bibnamefont {Baldomir}},\ and\ \bibinfo
  {author} {\bibfnamefont {A.}~\bibnamefont {Satoh}},\ }\href
  {https://doi.org/10.1039/C8CP05271H} {\bibfield  {journal} {\bibinfo
  {journal} {Physical Chemistry Chemical Physics}\ }\textbf {\bibinfo {volume}
  {20}},\ \bibinfo {pages} {30445} (\bibinfo {year} {2018})}\BibitemShut
  {NoStop}%
\bibitem [{\citenamefont {Guardia}\ \emph {et~al.}(2012)\citenamefont
  {Guardia}, \citenamefont {{Di Corato}}, \citenamefont {Lartigue},
  \citenamefont {Wilhelm}, \citenamefont {Espinosa}, \citenamefont
  {Garcia-Hernandez}, \citenamefont {Gazeau}, \citenamefont {Manna},\ and\
  \citenamefont {Pellegrino}}]{Guardia2012}%
  \BibitemOpen
  \bibfield  {author} {\bibinfo {author} {\bibfnamefont {P.}~\bibnamefont
  {Guardia}}, \bibinfo {author} {\bibfnamefont {R.}~\bibnamefont {{Di
  Corato}}}, \bibinfo {author} {\bibfnamefont {L.}~\bibnamefont {Lartigue}},
  \bibinfo {author} {\bibfnamefont {C.}~\bibnamefont {Wilhelm}}, \bibinfo
  {author} {\bibfnamefont {A.}~\bibnamefont {Espinosa}}, \bibinfo {author}
  {\bibfnamefont {M.}~\bibnamefont {Garcia-Hernandez}}, \bibinfo {author}
  {\bibfnamefont {F.}~\bibnamefont {Gazeau}}, \bibinfo {author} {\bibfnamefont
  {L.}~\bibnamefont {Manna}},\ and\ \bibinfo {author} {\bibfnamefont
  {T.}~\bibnamefont {Pellegrino}},\ }\href {https://doi.org/10.1021/nn2048137}
  {\bibfield  {journal} {\bibinfo  {journal} {ACS Nano}\ }\textbf {\bibinfo
  {volume} {6}},\ \bibinfo {pages} {3080} (\bibinfo {year} {2012})}\BibitemShut
  {NoStop}%
\bibitem [{\citenamefont {de~la Presa}\ \emph {et~al.}(2012)\citenamefont
  {de~la Presa}, \citenamefont {Luengo}, \citenamefont {Multigner},
  \citenamefont {Costo}, \citenamefont {Morales}, \citenamefont {Rivero},\ and\
  \citenamefont {Hernando}}]{DeLaPresa2012}%
  \BibitemOpen
  \bibfield  {author} {\bibinfo {author} {\bibfnamefont {P.}~\bibnamefont
  {de~la Presa}}, \bibinfo {author} {\bibfnamefont {Y.}~\bibnamefont {Luengo}},
  \bibinfo {author} {\bibfnamefont {M.}~\bibnamefont {Multigner}}, \bibinfo
  {author} {\bibfnamefont {R.}~\bibnamefont {Costo}}, \bibinfo {author}
  {\bibfnamefont {M.~P.}\ \bibnamefont {Morales}}, \bibinfo {author}
  {\bibfnamefont {G.}~\bibnamefont {Rivero}},\ and\ \bibinfo {author}
  {\bibfnamefont {A.}~\bibnamefont {Hernando}},\ }\href
  {https://doi.org/10.1021/jp310771p} {\bibfield  {journal} {\bibinfo
  {journal} {Journal of Physical Chemistry C}\ }\textbf {\bibinfo {volume}
  {116}},\ \bibinfo {pages} {25602} (\bibinfo {year} {2012})}\BibitemShut
  {NoStop}%
\bibitem [{\citenamefont {Brezovich}(1988)}]{Brezovich1988}%
  \BibitemOpen
  \bibfield  {author} {\bibinfo {author} {\bibfnamefont {I.}~\bibnamefont
  {Brezovich}},\ }\href@noop {} {\bibfield  {journal} {\bibinfo  {journal}
  {Med. Phys. Monogr.}\ }\textbf {\bibinfo {volume} {16}},\ \bibinfo {pages}
  {82} (\bibinfo {year} {1988})}\BibitemShut {NoStop}%
\bibitem [{\citenamefont {Fa{\'\i}lde}\ \emph {et~al.}(2024)\citenamefont
  {Fa{\'\i}lde}, \citenamefont {Ocampo-Zalvide}, \citenamefont {Serantes},\
  and\ \citenamefont {Iglesias}}]{Fraile2024}%
  \BibitemOpen
  \bibfield  {author} {\bibinfo {author} {\bibfnamefont {D.}~\bibnamefont
  {Fa{\'\i}lde}}, \bibinfo {author} {\bibfnamefont {V.}~\bibnamefont
  {Ocampo-Zalvide}}, \bibinfo {author} {\bibfnamefont {D.}~\bibnamefont
  {Serantes}},\ and\ \bibinfo {author} {\bibfnamefont {{\`O}.}~\bibnamefont
  {Iglesias}},\ }\href {https://doi.org/10.1039/D4NR02045F} {\bibfield
  {journal} {\bibinfo  {journal} {Nanoscale}\ }\textbf {\bibinfo {volume}
  {16}},\ \bibinfo {pages} {14319} (\bibinfo {year} {2024})}\BibitemShut
  {NoStop}%
\bibitem [{\citenamefont {Ruta}, \citenamefont {Chantrell},\ and\ \citenamefont
  {Hovorka}(2015)}]{Ruta2015}%
  \BibitemOpen
  \bibfield  {author} {\bibinfo {author} {\bibfnamefont {S.}~\bibnamefont
  {Ruta}}, \bibinfo {author} {\bibfnamefont {R.}~\bibnamefont {Chantrell}},\
  and\ \bibinfo {author} {\bibfnamefont {O.}~\bibnamefont {Hovorka}},\ }\href
  {http://dx.doi.org/10.1038/srep09090} {\bibfield  {journal} {\bibinfo
  {journal} {Sci. Rep.}\ }\textbf {\bibinfo {volume} {5}},\ \bibinfo {pages}
  {9090} (\bibinfo {year} {2015})}\BibitemShut {NoStop}%
\bibitem [{\citenamefont {D{\'e}jardin}, \citenamefont {Kachkachi},\ and\
  \citenamefont {Kalmykov}(2008)}]{Dejardin2008}%
  \BibitemOpen
  \bibfield  {author} {\bibinfo {author} {\bibfnamefont {P.-M.}\ \bibnamefont
  {D{\'e}jardin}}, \bibinfo {author} {\bibfnamefont {H.}~\bibnamefont
  {Kachkachi}},\ and\ \bibinfo {author} {\bibfnamefont {Yu.~P.}\ \bibnamefont
  {Kalmykov}},\ }\href {https://doi.org/10.1088/0022-3727/41/13/134004}
  {\bibfield  {journal} {\bibinfo  {journal} {J. Phys. D: Appl. Phys.}\ }\textbf
  {\bibinfo {volume} {41}},\ \bibinfo {pages} {134004} (\bibinfo {year} {2008})}%
  \BibitemShut {NoStop}%
\bibitem[{\citenamefont{Purushotham}\ and\ \citenamefont{Ramanujan}(2010)}]{PurushothamRamanujan2010}%
  \BibitemOpen
  \bibfield{author}{\bibinfo{author}{\bibfnamefont{S.}~\bibnamefont{Purushotham}}\ and\
  \bibinfo{author}{\bibfnamefont{R.~V.}\ \bibnamefont{Ramanujan}},\ }%
  \href{https://doi.org/10.1063/1.3432757}{\bibfield{journal}{\bibinfo{journal}{J.\ Appl.\ Phys.}\ }\textbf{\bibinfo{volume}{107}},\ \bibinfo{pages}{114701} (\bibinfo{year}{2010})}%
  \BibitemShut{NoStop}%
\bibitem[{\citenamefont{Narayanaswamy}\ \emph{et~al.}(2021)}]{Narayanaswamy2021}%
  \BibitemOpen
  \bibfield{author}{\bibinfo{author}{\bibfnamefont{V.}~\bibnamefont{Narayanaswamy}},
  \bibinfo{author}{\bibfnamefont{S.}~\bibnamefont{Sambasivam}},
  \bibinfo{author}{\bibfnamefont{A.}~\bibnamefont{Saj}},
  \bibinfo{author}{\bibfnamefont{S.}~\bibnamefont{Alaabed}},
  \bibinfo{author}{\bibfnamefont{B.}~\bibnamefont{Issa}},
  \bibinfo{author}{\bibfnamefont{I.~A.}\ \bibnamefont{Al-Omari}},\ and\
  \bibinfo{author}{\bibfnamefont{I.~M.}\ \bibnamefont{Obaidat}},\ }%
  \href{https://doi.org/10.3390/molecules26040796}{\bibfield{journal}{\bibinfo{journal}{Molecules}\ }\textbf{\bibinfo{volume}{26}},\ \bibinfo{pages}{796} (\bibinfo{year}{2021})}%
  \BibitemShut{NoStop}%
\bibitem[{\citenamefont{Michels}\ and\ \citenamefont{Kachkachi}(2026)}]{MichelsKachkachi2026}%
  \BibitemOpen
  \bibfield{author}{\bibinfo{author}{\bibfnamefont{A.}~\bibnamefont{Michels}}\ and\
  \bibinfo{author}{\bibfnamefont{H.}~\bibnamefont{Kachkachi}},\ }%
  \href{https://doi.org/10.1016/j.jmmm.2026.174461}{\bibfield{journal}{\bibinfo{journal}{J.\ Magn.\ Magn.\ Mater.}\ }\textbf{\bibinfo{volume}{656}},\ \bibinfo{pages}{174461} (\bibinfo{year}{2026})}%
  ; \href{https://doi.org/10.48550/arXiv.2603.15376}{arXiv:2603.15376}%
  \BibitemShut{NoStop}%
\end{thebibliography}

\end{document}